\documentstyle[preprint,prb,aps,epsf]{revtex}
\tightenlines
\begin{document}
\draft
\title{Plasma instability and amplification of electromagnetic waves \\in
low-dimensional electron systems}
\author{S. A. Mikhailov\cite{email}}
\address{Max-Planck Institut f\"ur Physik komplexer Systeme, N\"othnitzer Str. 38, D-01187 Dresden, Germany}
\date{\today}
\maketitle
\begin{abstract}
A general electrodynamic theory of a grating coupled two dimensional electron
system (2DES) is developed. The 2DES is treated quantum mechanically, the
grating is considered as a periodic system of thin metal strips or as an array
of quantum wires, and the interaction of collective (plasma) excitations in the
system with  
electromagnetic field is treated within the classical electrodynamics. It is 
assumed that a dc current flows in the 2DES. We consider a propagation of an 
electromagnetic wave through the structure, and obtain analytic dependencies of
the transmission, reflection, absorption and emission coefficients on the
frequency of light, drift velocity 
of 2D electrons, and other
physical and geometrical parameters of the system. If the drift velocity of 2D
electrons exceeds a threshold value, a 
current-driven plasma instability is developed in the system, and an incident
far infrared radiation is amplified. We show that in the
structure with a {\it quantum wire grating} the
threshold velocity of the amplification can be essentially reduced, as
compared to the commonly employed metal grating, down to experimentally
achievable values. Physically this is due to a considerable enhancement of the grating coupler
efficiency because of the resonant interaction of plasma modes in the 2DES
and in the grating. We show that tunable far
infrared emitters, amplifiers and generators can thus be created at realistic
parameters of modern semiconductor heterostructures.
\end{abstract}
\pacs{PACS numbers: 78.66.-w, 73.20.Mf, 42.55.Px, 78.45.+h}

\section{Introduction}

The motion of a fast electron beam across a periodic metal structure results in
the radiation of electromagnetic waves. This phenomenon, often referred 
to as the Smith-Purcell effect, \cite{SmithPurc} provides the basis for a
number of vacuum devices, such as travelling wave tubes and backward 
wave tubes. 
In these devices  electrons accelerated by an applied electric field up to
a velocity $v_{\rm dr}$, move in vacuum across a periodic metal structure
(a grating or a spirale), which leads to an amplification or generation of
electromagnetic waves at the frequency $f\sim v_{\rm dr}/a$, where $a$ is the
grating period. The drift velocity $v_{\rm dr}$ here is determined by the
applied electric voltage, so that these devices are 
{\it  voltage tunable} amplifiers and generators.

The vacuum devices successfully operate in the radio and microwave
range. A further enhancement of the operating frequency presents severe
difficulties because of the  
mechanical instability of a freely standing in vacuum periodic structure of
metal wires with a very small period. The operating
frequency of vacuum devices cannot therefore be extended up to the far
infrared (FIR) range.

In the late 70s an active experimental research of plasma oscillations in
two-dimensional (2D) electron systems (ES's) in Si MOSFETs
(metal-oxide-semiconductor field-effect transistors) and
GaAs/Al$_x$Ga$_{1-x}$As heterostructures has been
started. \cite{Allen77,Theis77,Tsui78,Theis78} In a considerable
part of experimental work (for a review see Ref.\ \onlinecite{Theis}) a FIR  
{\it transmission} spectroscopy has been used for the detection of 2D plasmons
(Fig.\ref{geometry}, where $v_{\rm dr}=0$). In this technique, the 2D plasmons
which are normally nonradiative modes, \cite{Stern} are coupled to
electromagnetic radiation by a metal grating placed in the vicinity of the 2D
layer. An incident electromagnetic wave with the intensity $I_0$ and the
electric field  polarized perpendicular to the grating strips, passes through
the structure in the direction perpendicular to the 2D layer ($z$ direction),
and the spectrum of the transmitted wave $T(\omega)$ is registered. Well
defined resonances which correspond to an excitation of 2D plasmons with
reciprocal lattice 
vectors ${\bf G}_m=(2\pi m/a,0)$ are observed in the transmission
spectrum (here $a$ is the grating period
and $m$ is integer). In experiments
of this type the energy of the external electromagnetic wave is converted to
the energy of the 2D plasmons. 

Figure \ref{geometry} resembles the geometry of the tunable vacuum amplifiers
and generators: the system consists in a conducting electron (2D) layer 
where electrons can move under the action of an applied electric field, and an
adjacent grating. First attempts to observe the emission of light from 
the grating coupled 2DES have been made in 1980. \cite{Tsui80,Gornik80} In
these FIR {\it emission} experiments (Fig.\ref{geometry}, where the drift
velocity $v_{\rm dr}\neq 0$, but the intensity of the incident light $I_0$ is
zero) a strong dc current is passed 
through the 2D layer (in the $x$ direction perpendicular to the grating
strips), and the emitted electromagnetic radiation is registered. The grating
period in these solid-state structures can be 
made smaller than 1 $\mu$m, and the typical frequency of 2D plasmons falls in
the terahertz (FIR) range. A successful realization of the 2D plasmon emission
experiment could lead to a creation of a tunable solid-state  
sourse of the FIR electromagnetic radiation. In spite of the strong
appeal of this idea and a number of more recent experiments,
\cite{Hopfel82PRL,Hopfel82,Okisu,Hirakawa95} the intensity of radiation from
the 2DES remains to be very small, and successfully working solid-state devices
based on the discussed principle are absent so far. 

The energy of the dc current passing through an electron system in the
presence of the grating is converted to electromagnetic radiation by two
steps. First, it is transformed to the energy of plasma oscillations in the
beam by means of a current-driven  plasma instability.
\cite{mikhailovsky,Chaplik80} Then the energy stored in the plasmon field
is converted to  electromagnetic radiation by means of the grating. The
plasma instability is developing in the system when the drift velocity of
electrons exceeds a threshold value $v_{\rm th}$, 
estimated as the plasma frequency of electrons in the beam  divided by a
typical grating wavevector $G_1=2\pi/a$ [see Eq. (\ref{threshveloc}) below, as well as
Ref.\ \onlinecite{Chaplik80}]. In vacuum devices the plasma frequency in the
electron beam is much smaller than that in the 2DES, and electrons can
be accelerated up to velocities much higher than those achievable in 
solid-state structures. The threshold condition for amplification is thus more
difficult to satisfy in solid-state structures than in their vacuum
counterparts, therefore the  
first attempts to realize the emission of light from the grating coupled 2DES
have failed. 

The aim of this paper is to develop a general theory of the transmission,
amplification and emission of light in the structure ``grating coupler
-- 2DES'', and to find realistic ways for a reduction of the threshold
velocity of amplification. We consider a propagation of light through
the structure ``grating coupler -- 2DES {\it with a flowing current}'' (Figure
\ref{geometry} with $v_{\rm dr}\neq 0$ and $I_0\neq 0$), and calculate the
transmission $T(\omega,v_{\rm dr})$, reflection $R(\omega,v_{\rm dr})$,
absorption $A(\omega,v_{\rm dr})$ and emission $E(\omega,v_{\rm dr})$ (at
$I_0=0$) 
coefficients as a function of the light frequency $\omega$, the drift velocity
$v_{\rm dr}$, as well as other physical and geometrical parameters of the
system. In the literature the problem of the transmission of light ($v_{\rm
dr}=0$, $I_0\neq 0$) has been solved analytically in a {\it
perturbative} approach \cite{Chaplik85,Theis} (the grating has been treated as
an infinitely thin metal layer with a weakly modulated density), and
numerically in a nonperturbative 
approach. \cite{Zheng} An emission of light from the structure ``metal grating
-- 2DES'' ($v_{\rm dr}\neq 0$, $I_0=0$) has been considered by Kempa {\it et
al.} \cite{Kempa93} using a numerical nonperturbative approach. We solve 
the problem  {\it analytically} using the {\it nonperturbative}
technique recently proposed in Refs.\ \onlinecite{SMdots,SMwires}. One of the
main results of our work is that the amplification of waves can be drastically
increased, and the threshold velocity can be essentially
reduced (down to experimentally achievable values) in structures ``2DES --
{\it quantum wire } grating'' (contrary to commonly employed structures with
{\it metal} gratings). The effect is due to the resonant interaction
of 2D plasmons with plasmons of the grating, which leads to a remarkable
enhancement of the grating coupler efficiency, and finally to an
improvement of device characteristics.

The paper is organized as follows. In Section \ref{formalism} we develop a
general theory of the scattering of electromagnetic waves on the
structure ``grating coupler -- thin conducting layer''. In Section
\ref{nodrift} we apply the general formalism to an analysis of the FIR
transmission, reflection and absorption 
spectra of the grating coupled 2DES without the dc electric current. 
In Section \ref{drift} we study an amplification of FIR radiation passing
through the system with a flowing current. In Section
\ref{emission} we discuss an emission spectrum of the structure ($v_{\rm dr}\neq 0$, $I_0= 0$) and compare
our approach with that of Ref.\ \onlinecite{Kempa93}. In Section
\ref{discussion} we summarize our results and formulate particular
recommendations for a designing  tunable solid-state FIR amplifiers. 

\section{scattering of light on a grating coupled thin conducting layer: General theory}
\label{formalism}

In this Section we develop a general theory of the scattering of light on a structure ``grating coupler -- thin conducting layer'' (Fig.\ref{geometry}). The grating coupler is
treated as an infinitely thin \cite{note0} conducting layer with an electron density 

\begin{equation}
N_1(x)\delta(z)=\sum_k n_1(x-ak)\delta(z),
\label{density}
\end{equation}

\noindent
placed in the plane $z=0$. The continuous function $n_1(x)$ is assumed to be
zero  at $|x|>W/2$ and an arbitrary nonzero function at $|x|<W/2$, where $W$
is the width of the grating strips and $a$ is the 
grating period. The conducting layer (2DES) is  placed in the plane $z=D$ and
described by the frequency and wave-vector dependent conductivity
$\sigma_{2D}({\bf q},\omega)$ [all quantities related to the grating (2DES) will
be supplied by the index 1 or 1D (2 or 2D)]. Electromagnetic wave is assumed
to be incident upon the structure
along the $z$ axis with the electric vector  being polarized in the 
$x$ direction, perpendicular to the grating strips. The system is infinite in the
$y$ direction, and a background dielectric constant $\epsilon$ is uniform in all the space. 

The total electric field ${\bf E}^{\rm tot}$ satisfies the Maxwell equations,

\begin{equation}
\nabla \times (\nabla \times {\bf E}^{\rm tot}) + \frac {\epsilon} {c^2}
\frac {\partial^2 {\bf E}^{\rm tot}} {\partial t^2} = - \frac {4\pi}{c^2}
\frac{\partial}{\partial t} \left[{\bf j}^{1D}(x)\delta(z)+{\bf j}^{2D}(x)\delta(z-D)\right],
\end{equation} 

\noindent
with scattering boundary conditions at $z\rightarrow\pm\infty$. We search for
a solution in the form

\begin{equation}
{\bf E}^{\rm tot}({\bf r},z)=\sum_{\bf G}\left[{\bf E}^{\rm
ext}_{\bf G}(z)+{\bf E}^{\rm ind}_{\bf G}(z)\right]e^{i{\bf G}\cdot{\bf
r}-i\omega t},
\end{equation}
where ${\bf G}={\bf G}_m=(2\pi m/a,0)$, and the  incident (external) and the
scattered (induced) electromagnetic waves are written as

\begin{equation}
{\bf E}^{\rm ext}_{\bf G}(z)=
\left(
\begin{array}{c}
E^{\rm ext}_{x,{\bf G}}\\
0\\0
\end{array}
\right)
e^{i\omega\sqrt{\epsilon}z/c}, 
{\ \ }E^{\rm ext}_{x,{\bf G}}=E_0\delta_{\bf G,0},
\end{equation}
and ${\bf E}^{\rm ind}_{\bf G}(z)=\left(
E^{\rm ind}_{x,{\bf G}}(z),
0,
E^{\rm ind}_{z,{\bf G}}(z)
\right)$. The field $E^{\rm ind}_{x,{\bf G}}(z)$ satisfies the equation 

\begin{equation}
\frac{\partial^2E_{x,{\bf G}}^{\rm ind}}{\partial z^2}-\kappa^2_{\bf
G}E_{x,{\bf G}}^{\rm ind}=\frac{4\pi i \kappa^2_{\bf
G}}{\omega\epsilon}[j^{1D}_{x,{\bf G}}\delta(z)+j^{2D}_{x,{\bf
G}}\delta(z-D)],
\label{DEforEx}
\end{equation}

\noindent
and has a solution

\begin{mathletters}
\begin{equation}
E_{x,{\bf G}}^{\rm ind}(z)|_{z<0}=
A_{\bf G}\exp(\kappa_Gz),
\end{equation} 

\begin{equation}
E_{x,{\bf G}}^{\rm ind}(z)|_{0<z<D}=
B_{\bf G}\sinh(\kappa_Gz)+C_{\bf G}\cosh(\kappa_Gz),
\end{equation} 

\begin{equation}
E_{x,{\bf G}}^{\rm ind}(z)|_{z>D}=
D_{\bf G}\exp(-\kappa_Gz),
\end{equation}
\label{induced}
\end{mathletters}
where
\begin{equation}
\kappa_{\bf G}\equiv\kappa_G=\sqrt{G^2-\omega^2\epsilon/c^2}. 
\end{equation}
If $G_m=2\pi m/a=0$, the value $\kappa_{G=0}=-i\omega\sqrt{\epsilon}/c$ is imaginary
(the radiative boundary conditions at 
$z\rightarrow\pm\infty$ imply that Im $\kappa_G<0$), and the values $A_{\bf
G=0}$ 
and $E_0+D_{\bf G=0}$ give the amplitudes of {\it
normally} reflected and transmitted waves. If $\kappa_G$ is imaginary
for several nonzero $m$ (i.e. at $\omega\sqrt{\epsilon}/c>G_m$), the  values $A_{\bf G}$ 
and $D_{\bf G}$ describe the amplitudes of
reflected and transmitted waves in corresponding ($m$-th) diffraction orders. 
For all $|m|>a/\lambda$, where $\lambda=2\pi
c/\omega\sqrt{\epsilon}$ is the wavelength of light, $A_{\bf G}$ 
and $D_{\bf G}$ give  the amplitudes of evanescent
(nonpropagating) electric field.

Using boundary conditions at the planes $z=0$ and $z=D$ we relate the
amplitudes of the  electric field $A_{\bf G},\dots,D_{\bf G}$ to the Fourier components of
the electric current:

\begin{equation}
A_{\bf G}=C_{\bf G}=-\frac{2\pi i\kappa_G}{\omega\epsilon}\left[j^{1D}_{x,{\bf
G}}+j^{2D}_{x,{\bf G}}\exp(-\kappa_GD)\right],
\label{A}
\end{equation}

\begin{equation}
B_{\bf G}=\frac{2\pi i\kappa_G}{\omega\epsilon}\left[j^{1D}_{x,{\bf
G}}-j^{2D}_{x,{\bf G}}\exp(-\kappa_GD)\right],
\label{B}
\end{equation}

\begin{equation}
D_{\bf G}=-\frac{2\pi i\kappa_G}{\omega\epsilon}\left[j^{1D}_{x,{\bf
G}}+j^{2D}_{x,{\bf G}}\exp(\kappa_GD)\right].
\label{D}
\end{equation}
Together with the relation

\begin{equation}
j^{2D}_{x,{\bf G}}=\sigma_{2D}({\bf G},\omega)E^{\rm tot}_{x,{\bf G}}|_{z=D},
\label{j2D}
\end{equation}
between the current in the 2DES and the  total
electric field at the plane $z=D$ we have five equations for six unknowns 
$A_{\bf G},\dots,D_{\bf G}$, $j^{1D}_{x,{\bf G}}$, $j^{2D}_{x,{\bf G}}$. Using
these equations we relate the total field at the plane $z=0$ to the current 
$j^{1D}_{x,{\bf G}}$ at the same plane,

\begin{equation}
E^{\rm tot}_{x,{\bf G}}|_{z=0}= W({\bf G},\omega)
\left( E^{\rm ext}_{x,{\bf G}}|_{z=0}-\frac{2\pi
i\kappa_G}{\omega\epsilon}j^{1D}_{x,{\bf G}}
\right),
\label{Etot}
\end{equation}
where 

\begin{equation}
W({\bf G},\omega)=
1-\left(1-\frac 1{\epsilon_{2D}({\bf
G},\omega)}\right)e^{-2\kappa_GD},
\label{Wfunc}
\end{equation}
and 

\begin{equation}
\epsilon_{2D}({\bf G},\omega)=1+\frac{2\pi
i\kappa_G}{\omega\epsilon}\sigma_{2D}({\bf G},\omega)
\label{dielfunc}
\end{equation}
is the (relative) ``dielectric permittivity'' of the 2DES. 

Properties of the grating should now be introduced into the theory. Usually 
\cite{Chaplik85,Zheng,Kempa93} one assumes the local Ohm's law for the
grating, $j_x^{1D}(x)=\sigma_{1D}(x,\omega)E_x^{\rm tot}(x,z=0)$, where the
conductivity $\sigma_{1D}(x,\omega)$ is proportional to the local 
electron density (\ref{density}). Then Eq. (\ref{Etot}) is rewritten in the
form of an integral equation

\begin{equation}
E_x^{\rm tot}(x)=E_0W({\bf 0},\omega)+\left(\frac{\partial^2}{\partial x^2}+
\frac{\omega^2\epsilon}{c^2} \right) \int\frac{dx^\prime}W\vartheta(x^\prime)L(x-x^\prime)E_x^{\rm
tot}(x^\prime), 
\label{INTEQ}
\end{equation}
where $E_x^{\rm tot}(x)\equiv E_x^{\rm tot}(x,z=0)$,
$\vartheta(x)=n_1(x)/\langle n_1(x)\rangle$ is a normalized electron 
density in a grating strip, the kernel $L(x-x^\prime)$ is defined as

\begin{equation}
L(x-x^\prime)=\frac{2\pi if \langle
\sigma_{1D}(\omega)\rangle}{\omega\epsilon} \sum_{\bf G}\frac{W({\bf
G},\omega)}{\kappa_G} e^{i{\bf G\cdot(r-r^\prime)}},
\label{kernel}
\end{equation}
$f=W/a$ is the geometrical ``filling factor'' of the grating, and  the
angular brackets mean the average over the area of a grating strip,
$\langle\dots\rangle=\int (\dots) dx/W$. 

A general scheme of solving Eq. (\ref{INTEQ}) is presented in Appendix
\ref{derivation}. Here we solve this equation approximately,
\cite{SMdots,SMwires,MS97} assuming that 
the total (and induced) electric field  
{\it inside} the strips is uniform, $E_x^{\rm inside}\equiv E_x^{\rm
tot}(|x|<W/2,z=0)=const$. This approximation works very well in a 
metal grating if the frequency of electromagnetic wave is small as compared to
the plasma frequency of the metal, and the width of the grating
strips is large as compared to the Thomas-Fermi
screening length. Under these conditions the electric field
inside the strips is completely screened and $E_x^{\rm inside}=const=0$. This
approximation is also valid in a quantum wire (quantum dot) grating at an
arbitrary frequency, if the wires (dots) are considered in an oblate cylinder 
\cite{aleiner} (oblate spheroid \cite{Allen83}) model. This follows from the
well known fact that an internal
electric field in an arbitrary ellipsoid is uniform if the external
one is uniform. \cite{ESS} This is also valid for wires or dots
 formed by a parabolic confining potential. \cite{shikin89} The validity of
the model has been recently checked using a number of
numerical approaches in Ref.\ \onlinecite{schaich}. It has been shown that the
model gives reliable results for experimentally measured (macroscopic) values
like for instance the transmission coefficient.  

Assuming that $E_x^{\rm inside}=const$, we get a relation between the total
electric field inside the grating strips and the external field (see Appendix \ref{derivation}). It has a form of a response equation

\begin{equation}
E_x^{\rm inside}
=\frac{E_0}{\zeta(\omega)}, 
\label{respeq}
\end{equation}
where the response function $\zeta(\omega)$ is given by

\begin{equation}
\zeta(\omega)= \frac 1{W({\bf 0},\omega)}\Biggl(1+\frac{2\pi if \langle\sigma_{1D}(\omega)\rangle}{\omega\epsilon} 
\sum_{\bf G}\kappa_G\alpha({\bf G})W({\bf G},\omega)
\Biggr).
\label{respfunc}
\end{equation}
The form-factor in Eq. (\ref{respfunc})

\begin{equation}
\alpha({\bf G})=|\langle \vartheta(x) e^{i {\bf G\cdot r}} \rangle |^2
\label{formfactor}
\end{equation}
is determined by  Fourier components of the equilibrium electron density in
the grating strips. 

The response equation (\ref{respeq}) and the response function
(\ref{respfunc}) are the main points of our theory. Having derived these
equations we can now calculate fields and currents in all
the space. In particular, for  Fourier amplitudes of electric fields at
$z<0$ and $z>D$ which describe the reflected and transmitted field in all
diffraction orders as well as the evanescent field we get

\begin{equation}
\frac{A_{\bf G}}{E_0}=-\delta_{\bf G,0}+ W({\bf G},\omega) 
\left(\delta_{\bf G,0}-\frac
{2\pi if \kappa_G\langle\sigma_{1D}(\omega)\rangle\langle\vartheta(x)e^{-i{\bf G\cdot r}}\rangle}{\omega\epsilon\zeta(\omega)}
\right),
\label{answerA}
\end{equation}

\begin{equation}
\frac{D_{\bf G}}{E_0}=-\delta_{\bf G,0}+\frac 1{\epsilon_{2D}({\bf G},\omega)} 
\left(\delta_{\bf G,0}-\frac
{2\pi if \kappa_G\langle\sigma_{1D}(\omega)\rangle\langle\vartheta(x)e^{-i{\bf G\cdot r}}\rangle}{\omega\epsilon\zeta(\omega)}
\right),
\label{answerD}
\end{equation}
Equations (\ref{answerA}), (\ref{answerD}) give the general solution of the
formulated problem. They have been derived under quite general assumptions and
include both the electrodynamics of the grating coupler and nonlocal and
quantum-mechanical effects in the response of the 2DES which enter the theory
via an appropriate model of the conductivity $\sigma_{2D}({\bf G},\omega)$. In
subsequent Sections we apply the general theory to the problem of FIR response
of the system ``grating coupler -- 2DES with and without a flowing current''.

\section{theory of the grating coupler: transmission of FIR radiation}
\label{nodrift}

\subsection{Approximations and preliminary notes}
\label{approx}

Before applying the general results of Section \ref{formalism} to the
problem of FIR response of the grating coupled 2DES we specify the conditions
of a typical experiment and make necessary approximations. First, we assume
that the grating
period $a$ is small as compared to the wavelength of light $\lambda$ (in a
typical experimental situation $\lambda\sim 300$ $\mu$m, $a\lesssim
1\mu$m). Under this condition only the ${\bf G=0}$ components of the electric
field describe outgoing waves, while all components with ${\bf G\neq 0}$ are
evanescent. The reflection $r(\omega)$ and transmission $t(\omega)$ amplitudes
are then determined by the coefficients $A_{\bf G=0}$ and 
$D_{\bf G=0}$ respectively, and we have

\begin{equation}
r(\omega)=-1+ W({\bf 0},\omega) \left(1-\frac
{2\pi f \langle\sigma_{1D}(\omega)\rangle}{c\sqrt{\epsilon}
\zeta(\omega)}\right),
\label{refl}
\end{equation}

\begin{equation}
t(\omega)=\frac 1{\epsilon_{2D}({\bf 0},\omega)}\left(1-\frac
{2\pi f \langle\sigma_{1D}(\omega)\rangle}{c\sqrt{\epsilon} \zeta(\omega)}\right),
\label{trans}
\end{equation}
where $\epsilon_{2D}({\bf 0},\omega)=1+2\pi\sigma_{2D}({\bf
0},\omega)/c\sqrt{\epsilon}$. 
The reflection, transmission and absorption coefficients are then determined,
as usual, by the relations \cite{note1}

\begin{equation}
R(\omega)=|r(\omega)|^2, {\ \ }T(\omega)=|t(\omega)|^2,
\label{RandT}
\end{equation}
\begin{equation}
A(\omega)=1-R(\omega)-T(\omega).
\label{absorption}
\end{equation}

Second, we assume that the distance $D$ between the 2DES and the grating is
also small as compared to $\lambda$. Then $
W({\bf 0},\omega)=\epsilon_{2D}^{-1}({\bf 0},\omega)$, 
and 
\begin{equation}
r(\omega)=-1+t(\omega),
\end{equation}
where $t(\omega)$ is given by Eq. (\ref{trans}).

Third, we specify the model for the conductivity of the 2DES. We postpone an
analysis of the nonlocal and quantum-mechanical effects in the 2DES to a
subsequent publication, and describe the properties of the 2DES in 
the hydrodynamic model \cite{Fetter} of $\sigma_{2D}({\bf q},\omega)$. 
Linearizing the continuity and Euler's equations \cite{note2} for the density
$n$ and the velocity ${\bf v}$ of 2D electrons,

\begin{equation}
\frac{\partial n}{\partial t} + {\bf \nabla}(n{\bf v})=0,
\end{equation}
\begin{equation}
\frac{\partial {\bf v}}{\partial t} + ({\bf v\cdot\nabla}){\bf v}=-\frac
e{m_2} 
{\bf E} - \gamma_2 ({\bf v-v}_{\rm dr}),
\label{euler}
\end{equation}
where ${\bf E}={\bf E}_0+\delta {\bf E}$, $n=n_2+\delta n$, ${\bf v}={\bf
v}_{\rm dr}+\delta {\bf v}$, we get ${\bf v}_{\rm dr}=-(e/m_2
\gamma_2){\bf E}_0$, and

\begin{equation}
\sigma_{2D}({\bf q},\omega)=\frac{n_2e^2}{m_2}\frac{i\omega}{(\omega-{\bf qv}_{\rm
dr})(\omega-{\bf qv}_{\rm dr}+i\gamma_2)}.
\label{conduct2}
\end{equation}
For the
average conductivity of the grating we assume, similarly, 

\begin{equation}
\langle\sigma_{1D}(\omega)\rangle=\frac{n_1e^2}{m_1}\frac i{\omega+i\gamma_1}.
\label{conduct1}
\end{equation}
In Eqs. (\ref{euler})--(\ref{conduct1}) $n_i$, $m_i$, and $\gamma_i$ are
the average electron density, the electron effective mass, and the  momentum
relaxation rate in the grating ($i=1$) and in the 2DES ($i=2$), respectively. 

Substituting the model expressions (\ref{conduct2}), (\ref{conduct1}) for the
conductivity of the 2DES and the grating into Eq. (\ref{trans}) we get the
following result for the transmission amplitude 
\cite{MS97}

\begin{equation}
t(\omega)={\omega+i\gamma_2 \over {\omega+i\gamma_2+i\Gamma_2}} 
\Biggl(1-{i\Gamma_1 \over {\omega+i\gamma_1}}{1 \over {\zeta (\omega)}}\Biggr),
\label{trans1}
\end{equation} 
where the response function assumes the form

\begin{eqnarray}
&&\zeta(\omega)=
\frac{i\Gamma_1}{\omega+i\gamma_1}+
\left(
1+\frac{i\Gamma_2}{\omega+i\gamma_2}
\right) \nonumber \\
\times && \left\{
1-\frac{\tilde\omega_{p1}^2}{\omega(\omega+i\gamma_1)}
\left[1+\frac{2\pi fn_1e^2}{m_1\epsilon\tilde\omega_{p1}^2} 
\sum_{m\neq 0}
\frac{|G_m| \alpha(G_m)e^{-2|G_m|D} \omega_{p2}^2(G_m)}
{(\omega-G_mv_{\rm dr}) (\omega-G_mv_{\rm dr}+i\gamma_2) -\omega_{p2}^2(G_m) }
\right]
\right\}.
\label{zeta1}
\end{eqnarray} 
Here
\begin{equation}
\omega_{p2}(G_m)=\left(\frac{2\pi n_2e^2|G_m|}{m_2\epsilon}\right)^{1/2}
\label{2Dplasmons}
\end{equation} 
is the frequency of 2D plasmons, $G_m=2\pi m/a$, 

\begin{equation}
\Gamma_1=\frac{2\pi fn_1e^2}{m_1c\sqrt{\epsilon}},
\label{rad1}
\end{equation}

\begin{equation}
\Gamma_2=\frac{2\pi n_2 e^2}{m_2c \sqrt{\epsilon}},
\label{rad2}
\end{equation}
and 

\begin{equation}
\tilde\omega_{p1}^2 = \frac{2\pi f n_1e^2}{m_1\epsilon} \sum_{m\neq 0}
|G_m| \alpha(G_m).
\label{1Darray}
\end{equation} 
The physical meaning of the values $\Gamma_1$, $\Gamma_2$, and
$\tilde\omega_{p1}$ will be discussed in 
the Section \ref{simplecases}. 

Finally, we specify the density profile function $\vartheta(x)$ which
determines the form-factor $\alpha({\bf G})$. In
the main part of the paper we will use a semielliptic density profile, 

\begin{equation}
\vartheta(x)=\case 4\pi[1-(2x/W)^2]^{1/2},
\label{elliptic}
\end{equation} 
for which the form-factor $\alpha(G_m)$ is given by

\begin{equation}
\alpha(G_m)=[2J_1(z)/z]^2, {\ \ }z=G_mW/2,
\label{ffelliptic}
\end{equation} 
where $J_1$ in the Bessel function. In some cases we will also
consider a step-like profile, $\vartheta_{\rm step}(x)=\theta(W/2-|x|)$, for which

\begin{equation}
\alpha_{\rm step}(G_m)=(\sin z/z)^2, {\ \ }z=G_mW/2.
\label{ffstep}
\end{equation} 

Using equations (\ref{trans1}), (\ref{zeta1}), as well as (\ref{RandT}) and
(\ref{absorption}) one can show that the 
functions $T(\omega)$, $R(\omega)$ and $A(\omega)$ have one or more resonant
features related with an excitation of plasma modes in the system. The
resonance frequencies and linewidths depend on the drift velocity
of 2D electrons $v_{\rm dr}$, as well as on other physical and geometrical
parameters of the 
structure. If a resonant feature is well separated from others,
i.e. when its linewidth is small as compared to the distance between adjacent
resonances, the functions $T(\omega)$, $R(\omega)$ and $A(\omega)$ assume the
following general form

\begin{mathletters}
\begin{equation}
T(\omega)=1-\frac{\omega^2(2\gamma
\Gamma
+\Gamma
^2)}
{(\omega^2-\Omega^2)^2+\omega^2(\gamma
+\Gamma
)^2},
\end{equation} 

\begin{equation}
R(\omega)=\frac{\omega^2\Gamma
^2}
{(\omega^2-\Omega^2)^2+\omega^2(\gamma
+\Gamma
)^2},
\end{equation}

\begin{equation}
A(\omega)=\frac{2\omega^2\gamma
\Gamma
}
{(\omega^2-\Omega^2)^2+\omega^2(\gamma
+\Gamma
)^2},
\end{equation}
\label{TRAgeneral}
\end{mathletters}
\noindent
where $\Omega$ is the resonance frequency, $\gamma
$ and $\Gamma
$ are the nonradiative and the radiative decay rates, respectively. The
total linewidth of the resonance is thus determined by the sum of the
radiative and the nonradiative decay rates, the resonant values of the
transmission, reflection and absorption coefficients, $T_{\rm res}\equiv
T(\Omega)$, $R_{\rm res}\equiv R(\Omega)$, $A_{\rm res}\equiv A(\Omega)$,
are determined by the ratio $\gamma
/\Gamma
$,

\begin{equation}
T_{\rm res}=\frac{\gamma^2}{(\gamma+\Gamma)^2}, {\ } 
R_{\rm res}=\frac{\Gamma^2}{(\gamma+\Gamma)^2}, {\ }
A_{\rm res}=\frac{2\gamma\Gamma}{(\gamma+\Gamma)^2}.
\label{resvalues}
\end{equation}
The resonant values (\ref{resvalues}) characterize the strength of the
resonant features. Note that the reflection (absorption) resonant amplitude is
negligible as compared to the absorption (reflection) amplitude if
$\Gamma
\ll\gamma
$ ($\Gamma
\gg\gamma
$). In
the following, we  specify the values of $\Omega$, $\gamma
$ and
$\Gamma
$ in different considered cases.

The rest of this Section is devoted to an analysis of the system without
flowing current, i.e. at $v_{\rm dr}=0$. In Section \ref{drift} we analyze the
general formulas at a finite drift velocity. 

\subsection{Two limiting cases}
\label{simplecases}

We start our analysis from two simple limiting cases, of a 2DES without grating
coupler and of a grating without the 2DES.
 
\subsubsection{2DES without  grating}
\label{2DESwithout}

If the grating is absent, then $\Gamma_1=0$, and the transmission,
reflection and absorption coefficients assume the form (\ref{TRAgeneral}) with

\begin{equation}
\Omega=0, {\ \ }\gamma
=\gamma_2, {\ \ }\Gamma
=\Gamma_2.
\label{2DESparameters}
\end{equation}
The nonradiative contribution to the linewidth $\gamma
$ is determined
by the momentum relaxation rate of 2D electrons $\gamma_2$ and is due to the
Drude absorption in the 2DES. The physical meaning of the value 
$\Gamma_2$, Eq. (\ref{rad2}), is the radiative
decay of oscillating 2D electrons in the 2DES taken in isolation (without the
grating). Indeed, if one electron is placed in an   
electric field $E_0\exp(-i\omega t)$, it oscillates with an amplitude
$\delta x\sim e E_0/m\omega^2$. This creates an oscillating dipole moment
$d\sim e^2 E_0/m\omega^2$, which produces a dipole radiation
\cite{LandauTF} with the intensity $I\sim\omega^4d^2/c^3$. Dividing the
radiated intensity $I$ by the average energy of the oscillating dipole
$W\sim m\omega^2 \delta x^2$ one gets the radiative decay of a {\it single}
electron $\Gamma_0\sim e^2\omega^2/mc^3$. When a {\it sheet of electrons} with
an area density $n_s$ is placed in an oscillating electric field, and the
inter-electron distance $n_s^{-1}$ is small as compared to the wavelength of
light $\lambda$, all $N\sim n_s\lambda^2$ electrons within the coherence area
$\sim \lambda\times\lambda$ radiate in phase. The average energy should then
be multiplied by a factor of $N$, while
the radiated intensity by a factor of $N^2$. The radiative decay of
an electron sheet is then given by the product
$\Gamma_0N\sim\Gamma_0n_s\lambda^2\sim n_se^2/mc$ in agreement with the exact
expression (\ref{rad2}).

\subsubsection{Grating without 2DES}

If the 2DES is absent, then $\Gamma_2=\omega_{p2}(G_m)=0$, and the
transmission, reflection and absorption coefficients assume the form
(\ref{TRAgeneral}) with 

\begin{equation}
\Omega=\tilde\omega_{p1}, {\ \ } \gamma
=\gamma_1, {\ \ }\Gamma
=\Gamma_1. 
\label{Gratparameters}
\end{equation}
The nonradiative contribution to the linewidth $\gamma_1$ is now due to the
Drude absorption in the grating strips. The value $\Gamma_1$ is
proportional to the average electron density in the grating $fn_1$,
Eq. (\ref{rad1}), and is the radiative decay of plasma oscillations in the
grating taken in isolation (with the removed 2DES). The value
$\tilde\omega_{p1}$ gives the resonance frequency 
of plasmons in a periodic array of grating 
strips (or quantum wires). Eq. (\ref{1Darray}) provides a functional
dependence of 
$\tilde\omega_{p1}$ on the equilibrium electron density $\vartheta(r)$ in
wires (similar functional dependencies for arrays of quantum dots and antidots
have been found in Refs.\ \onlinecite{SMdots} and \onlinecite{SMantidots}
respectively). If the equilibrium electron density in strips has a
semielliptic form (\ref{elliptic}), Eq. (\ref{1Darray}) gives (see Appendix
\ref{wires}) 

\begin{equation}
\tilde\omega_{p1}^2 = \omega_{p1}^2\beta(f) \equiv \omega_{p1}^2 \left[1 -
\frac{(\pi f)^2}{24} - \frac{(\pi f)^4}{960}\right], 
\label{1Dresfreq}
\end{equation} 
where 

\begin{equation}
\omega_{p1}=\sqrt{\frac{16n_1e^2}{m_1\epsilon W}}
\label{singlewire}
\end{equation} 
is the resonance frequency of plasmons in a single wire, and the factor
$\beta(f)$ is due to the inter-wire interaction. If the wire is formed by an
external {\it parabolic} confining potential $V_{\rm ext}(x)=Kx^2/2$,
Eq. (\ref{1Darray}) reproduces an exact result $\omega_{p1}^2=K/m_1$ of the
generalized Kohn theorem, \cite{brey} see Appendix \ref{wires}.

Figure \ref{qwgratingonly} shows the frequency dependencies of the
transmission, reflection and absorption coefficients of a quantum wire array
at two ratios of the collisional damping to the radiative decay,
$\gamma_1/\Gamma_1\gg 1$ (Fig. \ref{qwgratingonly}a) and
$\gamma_1/\Gamma_1\ll 1$ (Fig. \ref{qwgratingonly}b). In the former case the
reflection of waves is negligibly small, and the transmission minimum is due
to a peak in the absorption coefficient. In the latter case,
the absorption of waves is small as compared to their reflection, and the
transmission minimum is mainly due to the reflection peak. The 
width of the resonance in the second case is smaller than that in the first
case (but does not tend to zero) and is determined mainly by the radiative
decay. 

\subsection{Grating coupled 2DES}

Now we consider the transmission of FIR radiation through a coupled structure
``grating -- 2DES'', under the condition when no current is flowing in the
2DES ($v_{\rm dr}=0$). We consider two different cases: the case of a {\it
metal} grating, when the plasma frequency in the grating $\tilde\omega_{p1}$
is much larger (several orders of magnitude in a typical experiment) than the
2D plasmon frequency $\omega_{p2}(G_1)$, and the case of a {\it
quantum-wire} grating, when the plasma frequencies in the grating and in the
2DES are of the same order of magnitude.

\subsubsection{Metal grating}
\label{SectionMetGr}

Figure \ref{MG_nodrift} demonstrates the frequency dependent transmission
coefficient of the structure ``metal grating -- 2DES'' at different values of the
geometrical filling factor of the grating $f$. Three characteristic features
are seen in the Figure. First, the position of resonances which we
denote as $\Omega_{12}(m)$  does not coincide
with the frequencies of the 2D plasmons $\omega_{p2}(G_m)$, 
Eq. (\ref{2Dplasmons}), shown in Figure \ref{MG_nodrift} by triangles for
four lowest harmonics $m=1,2,3,4$. The index `12' here reminds that we deal
with the {\it 
coupled} 1D (grating) $-$ 2D electron system. Second, the position and the
amplitude of resonances for different $m$ essentially depend on the filling
factor $f=W/a$. At some values of $f$ the amplitude of higher harmonics
can be comparable with or even larger than those  of lower harmonics
(as is the case for the modes $m=1$ and $m=2$ at $W/a=0.2$, Figure
\ref{MG_nodrift}c, or for the modes $m=2$ and $m=3$ at $W/a=0.6$, Figure
\ref{MG_nodrift}a). At certain values of $f$ some harmonics are not excited at
all (e.g., the mode $m=2$ at $W/a=0.6$, Figure 
\ref{MG_nodrift}a). Third, the amplitudes of the transmission resonances
become smaller when the resonance positions $\Omega_{12}(m)$ approach the 2D
plasmon frequencies $\omega_{p2}(G_m)$, see, e.g., the evolution of the
$\Omega_{12}(1)$ mode amplitude with decreasing $W/a$ (note the
difference in the vertical axis scales in Figures \ref{MG_nodrift}a-c). 

In order to understand these features we take the limits $v_{\rm dr}=0$ (no
drift) and $n_1\rightarrow\infty$ (metal grating) in the general formulas
(\ref{trans1}) -- (\ref{zeta1}). The resulting expressions for the
transmission, reflection and absorption 
coefficients near the resonance $\omega=\Omega_{12}(m)$ assume the form
(\ref{TRAgeneral}) where

\begin{equation}
\Omega^2=\Omega_{12}^2(m)\equiv\omega_{p2}^2(G_m)(1-\Delta_m),
\label{resfreq2DESmetgr}
\end{equation} 
\begin{equation}
\Gamma
=\Gamma_{12}(m)\equiv\frac{\Gamma_1}{\tilde\omega_{p1}^2}
\omega_{p2}^2(G_m)\Delta_m=\Gamma_2 \frac{(\pi f)^2}{4\beta(f)}|m|\Delta_m,  
\label{raddecay2DESmetgr}
\end{equation} 
and $\gamma
=\gamma_2$. 
The parameter

\begin{equation}
\Delta_m=\frac{(\pi f)^2}{2\beta(f)}|m|\alpha(G_m)\exp(-2|G_m|D)
\label{Delta}
\end{equation}
here depends on the harmonic index $m$ and on geometrical parameters of the
structure. Note that the physical parameters of the grating -- the electron
density $n_1$, the momentum relaxation rate $\gamma_1$, and the effective mass
$m_1$ -- do not enter the formulas (\ref{resfreq2DESmetgr}) -- (\ref{Delta}),
in which the grating is presented only via the geometrical parameters $a$,
$W$, and $D$. The resonant values of the transmission, reflection, and
absorption coefficients, $T_{\rm res}(m)$, $R_{\rm res}(m)$, and  $A_{\rm
res}(m)$, are determined by Eq. (\ref{resvalues}) where $\gamma=\gamma_2$ and
$\Gamma=\Gamma_{12}(m)$. 

As seen from Eqs. (\ref{resfreq2DESmetgr}), (\ref{raddecay2DESmetgr}), and
(\ref{resvalues}), the resonance frequency, the radiative contribution to
the linewidth, and the strength of the resonance essentially depend on the
parameter $\Delta_m$, which exponentially decreases with the distance $D$
between the 2DES and the grating and oscillates as a function of the grating
filling 
factor $f$ [via the oscillating $f$-dependence of the form-factor
$\alpha(G_m)$, Eqs. (\ref{ffelliptic}), 
(\ref{ffstep})]. If $\Delta_m$ tends to zero, the resonance frequencies
$\Omega_{12}(m)$ tend to the 2D plasmon frequencies $\omega_{p2}(G_m)$,
but the radiative decay and the 
strength of the resonances vanish, $\Gamma_{12}(m)=0$, $T_{\rm
res}(m)=1$. The value of $\Delta_m$ vanishes when $\pi fm$ coincides with any
of zeros of the Bessel function $J_1$, for a semielliptic density
profile (\ref{elliptic}), or with any of zeros of the sine function, for a step-like
profile. In order to get the maximum strength of the $m$-th plasma resonance
one should thus satisfy the condition $J_1^2(\pi fm)$=maximum (a semielliptic
profile), or

\begin{equation}
\frac Wam=0.57, {\ \ } \frac Wam=1.7, 
\label{bestconditions}
\end{equation}
etc. Figures \ref{frequencyfigure} and \ref{Tminfigure}
demonstrate the $f$-dependencies of the resonance frequencies $\Omega_{12}(m)$ [normalized by $\omega_{p2}(G_1)$], radiative contribution
to the linewidth $\Gamma_{12}(m)$ (normalized by $\Gamma_2$), and the
resonant transmission coefficient $T_{\rm res}(m)$ for three lowest modes
$m=1$, 2, and 3  at $D/a=0.08$, for the semielliptic density profile
(\ref{elliptic}). The 
behaviour of $\Omega_{12}(m)$, $\Gamma_{12}(m)$ and $T_{\rm res}(m)$
at the step-like profile is qualitatively the same.

\subsubsection{Quantum-wire grating}
\label{nodriftQW}

If the grating coupler is made out of a metal, only the geometrical (but not
the physical) parameters of the grating determine the observable transmission
(reflection, absorption) resonances. If the grating is made out of a 2D
electron layer with {\it similar} plasma parameters (the quantum-wire grating)
the observable resonances are determined by both the 2D plasmons in the 2DES,
and the plasma modes in the grating. This gives additional possibilities to
control the transmission spectra, especially in the finite drift velocity
regime (Section \ref{drift}). 

Figure \ref{QW_nodrift}  shows the transmission coefficient $T(\omega)$ of the
structure ``{\it quantum-wire} grating -- 2DES'' at three different values of
the grating plasma frequency $\tilde\omega_{p1}$. Geometrical parameters of
the structure are the same as in Fig.\ \ref{MG_nodrift}c, where the
transmission coefficient of the {\it metal} grating coupled 2DES is
shown. Three new features are seen in Figure \ref{QW_nodrift} as compared to
Figure \ref{MG_nodrift}c. First, due to the 
presence of the grating plasmon resonance $\tilde\omega_{p1}$ an additional
resonance peak appears in the plot. Second, due to the interaction of 2D
plasmons and the grating plasmon the resonance peaks are slightly
shifted relative to their positions in Fig.\ \ref{MG_nodrift}c. Third, and the
most important feature  is a dramatic
enhancement of the amplitudes of the 2D plasmon resonances, in situations when
the frequency $\tilde\omega_{p1}$ approaches the 2D plasmon frequencies, Figure
\ref{QW_nodrift}a,b. This effect is due to a resonant
interaction of the grating plasmon with the 2D ones, and is especially
pronounced for higher 2D plasmon modes, for which the amplitude of resonances
is increased by about an order of magnitude (note the difference in the
vertical axis scales in Figures \ref{QW_nodrift} and \ref{MG_nodrift}c). This
effect is of a particular importance in the 2DES with a flowing current, as it
allows one to increase the amplification and to reduce the threshold velocity
in the structure with the quantum wire grating (Section \ref{driftQW}). 

\section{amplification of waves}
\label{drift}

Now we analyze the transmission of electromagnetic waves through a grating
coupled 2DES with a flowing current. We start from the case of a
structure ``metal grating -- 2DES''. 

\subsection{Metal grating}
\label{driftMG}

Figure \ref{MG_drift} shows the absorption coefficient of the structure
``metal grating -- 2DES'' at relatively small values of the dc current
($U\equiv v_{\rm dr}/v_{F2}\leq 1.6$) in the frequency interval corresponding
to an excitation of the first ($m=1$) 2D plasmon harmonic. Here $v_{F2}$ is
the Fermi velocity of 2D electrons in the 2DES. The physical and
geometrical parameters of the structure are the same as in Figure
\ref{MG_nodrift}a ($W/a=0.6$). The triangle at the bottom of the plot shows
the position of the $m=1$ 2D plasmon harmonic (\ref{2Dplasmons}). As seen from
the Figure \ref{MG_drift}, in contrast to the case of the vanishing current
($U=0$, lower curve), at finite drift velocities of 2D electrons there are two
modes, $\Omega_{12}(m,\pm)$, associated with each harmonic number
$m$. We label these modes by two indexes ($m,\pm$), where the frequency of the
$+$ ($-$) mode increases (decreases)
with $v_{\rm dr}$ at small $v_{\rm dr}$, $d\Omega_{12}(m,+)/dv_{\rm dr}>0$ and
$d\Omega_{12}(m,-)/dv_{\rm dr}<0$ at $v_{\rm dr}\rightarrow 0$. As seen from
the Figure, 

\begin{equation}
\lim_{v_{\rm dr}\rightarrow 0} \Omega_{12}(m,+)=\omega_{p2}(G_m),
\end{equation}

\begin{equation}
\lim_{v_{\rm dr}\rightarrow 0} \Omega_{12}(m,-)=\Omega_{12}(m),
\end{equation}
where $\Omega_{12}(m)$ is defined in Eq. (\ref{resfreq2DESmetgr}). The
strength of the $(m,+)$ mode vanishes when the drift velocity tends to zero. 

In order to get a quantitative description of the resonant features shown on
Figure \ref{MG_drift} we take the limit $n_1\rightarrow\infty$ (metal grating)
in the general formulas (\ref{trans1}) -- (\ref{zeta1}). Assuming that 
$|\omega+i\gamma_2|\gg\Gamma_2$ and $\omega_{p2}(G_m)\gg\gamma_2$, and taking
into account only the terms with $m=\pm|m|$ in the sum in Eq. (\ref{zeta1}),
we find that 
near the $(m,\pm)$ resonance the transmission, reflection and absorption
coefficients assume the form (\ref{TRAgeneral}), where the
resonance frequency $\Omega=\Omega_{12}(m,\pm)$ is determined by the equation

\begin{eqnarray}
\Omega_{12}^2(m,\pm)&&= \omega_{p2}^2(G_m)\left(1-\frac{\Delta_m}2
\right) + (G_mv_{\rm dr})^2 \nonumber \\
&&\pm \omega_{p2}(G_m) \sqrt{\omega_{p2}^2(G_m)\left(\frac{\Delta_m}2
\right)^2+(2G_mv_{\rm dr})^2 \left(1-\frac{\Delta_m}2
\right)},
\label{resfreqdrifted}
\end{eqnarray}
the radiative decay $\Gamma=\Gamma_{12}(m,\pm)$ is given by

\begin{equation}
\Gamma_{12}(m,\pm)= \frac{\Gamma_{12}(m)}2 \left\{
1 \mp \frac
{\omega_{p2}^2(G_m) \Delta_m - (2G_mv_{\rm dr})^2}
{\omega_{p2}(G_m)\sqrt{\omega_{p2}^2(G_m) \Delta_m^2 + (4G_mv_{\rm dr})^2 (1 -
\Delta_m/2)}}
\right\},
\label{raddecaydrifted}
\end{equation}
and the nonradiative decay $\gamma=\gamma_2$. Figure \ref{rfdec_drift} shows
the drift velocity dependencies of the resonance frequency
$\Omega_{12}(m,\pm)$ and the normalized radiative decay
$\Gamma_{12}(m,\pm)/\Gamma_{12}(m)$ for the mode $m=1$ at parameters of Figure
\ref{MG_nodrift}a. The frequency $\Omega_{12}(m,+)$ [$\Omega_{12}(m,-)$]
increases (decreases) with the square
of the drift velocity at $G_mv_{\rm dr}\ll\omega_{p2}(G_m)\Delta_m/4$, and
linearly, 

\begin{equation}
\Omega_{12}(m,\pm)\approx |\omega_{p2}(G_m)\sqrt{1-\Delta_m/2}\pm G_mv_{\rm
dr}|,
\label{Dopplershifted}
\end{equation}
at $G_mv_{\rm dr}\gg\omega_{p2}(G_m)\Delta_m/4$ (the Doppler shifted plasma
resonances). In the region 
$\omega_{p2}(G_m)\sqrt{1-\Delta_m}<G_mv_{\rm dr}<\omega_{p2}(G_m)$ the
frequency  $\Omega_{12}(m,-)$ vanishes, and at $G_mv_{\rm
dr}>\omega_{p2}(G_m)$ it increases again with the positive slope,
$d\Omega_{12}(m,-)/dv_{\rm dr}>0$ at $G_mv_{\rm
dr}>\omega_{p2}(G_m)$. \cite{note3} The radiative decay, as well as the
strength of the 
$\Omega_{12}(m,+)$ resonance, equal zero at $v_{\rm dr}=0$ and
increase monotonously when the drift velocity increases (Figure
\ref{rfdec_drift}b). The radiative decay of the $\Omega_{12}(m,-)$ mode
decreases from a finite value at $v_{\rm dr}=0$, vanishes at $G_mv_{\rm
dr}=\omega_{p2}(G_m)$, and {\it changes its sign} at $G_mv_{\rm
dr}>\omega_{p2}(G_m)$. As seen from Eqs. (\ref{resvalues}), when $\Gamma$
equals zero, the reflection and absorption coefficients at the resonance
$\omega=\Omega$ disappear, and the transmission coefficient equals unity. When
$\Gamma$ becomes negative, the resonant transmission coefficient exceeds
unity, which means an amplification of waves, while the absorption coefficient
becomes negative, which 
means that the energy is transfered not from the electromagnetic wave to
the electron system, but conversely, from the current driven electron system
(eventually from the battery which supplies the current) to the
electromagnetic wave. The plasma mode $\Omega_{12}(m,-)$ thus becomes unstable
at 
$G_mv_{\rm dr}>\omega_{p2}(G_m)$. \cite{Chaplik80} Figure \ref{TRAres_drift}
demonstrates the drift 
velocity dependencies of the transmission, reflection and absorption
coefficients at the resonance $\omega=\Omega_{12}(m,\pm)$ for
$m=1$. An amplification of the  
transmitted electromagnetic waves is explicitly demonstrated in Figure
\ref{MG_drift2} where we draw the frequency dependence of the transmission
coefficient at different drift velocities at a larger range (as compared with
Figure \ref{MG_drift}) of $v_{\rm dr}$ ($0\leq U\leq 8$). 

In the above calculations we have assumed that the resonance at
$\omega=\Omega_{12}(m,\pm)$ is well separated from other resonances (i.e. that
the width of the resonance line is small as 
compared to the distance to neighbour resonances). There are two effects
in which this approximation is insufficient. The first one concerns an
accurate evaluation of the threshold velocity of the amplification of waves. 
We define the threshold velocity $v_{\rm th}$  by the condition $T_{\rm
res}>1$. As follows from the above discussion and Figure \ref{TRAres_drift},
the resonant transmission coefficient exceeds unity when $G_mv_{\rm
dr}>\omega_{p2}(G_m)$. This inequality gives however only the lower estimate
for the $v_{\rm th}$. In order to evaluate the threshold
velocity more accurately one should take into account that the amplification
of waves should 
exceed the Drude absorption in the 2DES (subsection \ref{2DESwithout}) which
is essential at low frequencies. Including this fact we get 
the following expression for the threshold velocity

\begin{equation}
v_{\rm th}=\frac{\omega_{p2}(G_m)}{G_m}\left(1+\frac{\Delta_m}2X\right),
\label{threshveloc}
\end{equation}
where the (positive) factor $X$ is determined by the cubic equation

\begin{equation}
X^3+X^2=A\equiv
\frac{8\gamma_2\Gamma_2(2\gamma_2+\Gamma_2)}{\Gamma_{12}(m) 
\omega_{p2}^2(G_m)\Delta_m^3}.
\label{cubiceq}
\end{equation}
The threshold velocity thus consists in two contributions. The first one, 

\begin{equation}
\frac{\omega_{p2}(G_m)}{G_m}=v_{F2}\sqrt{\frac a{2\pi a_B^\star|m|}},
\label{1stcontrib}
\end{equation}
can be reduced by chosing the structures with a low 2D electron density $n_2$
and a small 
period $a$, and exploiting an excitation of higher 2D plasmon harmonics (here
$a_B^\star$ is the effective Bohr radius). The second contribution due to the
correction $\Delta_mX/2$ in Eq. (\ref{threshveloc}) has a complicated
dependence on the density $n_2$, momentum relaxation rate $\gamma_2$, the mode
index $m$ and the geometrical parameters of the structure. Qualitatively these
dependencies can be understood if to note that the factor $X$ 
 equals  $A^{1/3}$, if $A\gg 1$, and $A^{1/2}$, if $A\ll 1$, while the  
parameter $A$, in its turn, is proportional to

\begin{equation}
A\propto\gamma_2\left(1+\frac{2\gamma_2}{\Gamma_2}\right)
\frac{a \exp(8G_mD)}{f^2J_1^2(\pi fm)}
\label{2ndcontrib}
\end{equation}
(we consider the semielliptic density profile). Thus, the second contribution
to the threshold velocity can be reduced if the parameter $W/a$
satisfies the conditions (\ref{bestconditions}), the distance between the 2DES
and the grating is small as compared to the width of the grating strips, $D\ll
W$, and the grating period $a$, as well as the
momentum relaxation rate $\gamma_2$, are taken to be as small as possible. The
$f$ dependence of the normalized 
threshold velocity $v_{\rm th}/v_{F2}$ for several lowest modes $m=1,\dots,4$,
and  for parameters of Figure \ref{MG_nodrift} ($n_{s2}=3\times 10^{11}$
cm$^{-2}$, $\gamma_2=0.7\times 10^{11}$ s$^{-1}$, $D=60$ nm) is shown in 
Figure \ref{threshvel}. The divergencies of $v_{\rm th}$ are related to zeros
of the factor $\Delta_m$.

The second effect which is not described by our single-resonance approximation
is an anticrossing of modes with different $m$, which can be seen in
Figure \ref{MG_drift} at $U\approx 0.8$ and at $U\approx 1.5$, where the
relatively weak mode $(3,-)$
intersects the modes $(1,+)$ and $(1,-)$ respectively, as well as in Figure
\ref{MG_drift2} at $U\approx 5.0$, where the mode $(3,-)$ (which has a positive
slope with respect to $v_{\rm dr}$ at so large $U$) intersects the mode
$(1,+)$ for a second time. The intersection points of modes $(m_1,-)$ and
$(m_2,\pm)$ ($m_1>m_2$) are determined by the relation
$\Omega_{12}(m_1,-)=\Omega_{12}(m_2,\pm)$ [the modes $\Omega_{12}(m_1,+)$ and
$\Omega_{12}(m_2,\pm)$ do not intersect at $m_1>m_2$], and  the
transmission, reflection and absorption coefficients near the anticrossing can
be found from  Eqs. (\ref{trans1}) -- (\ref{zeta1}) in the
limit $n_1\rightarrow\infty$ (metal grating) if to take into account only the
terms with $m=m_1$ and $m=m_2$ in the sum in Eq. (\ref{zeta1}). The most
interesting situation is realized at a large drift velocity, $v_{\rm
dr}>\omega_{p2}(G_{m_1})/G_{m_1}$, when an {\it unstable} plasma mode
$\Omega_{12}(m_1,-)$ intersects one of the stable plasma modes
$\Omega_{12}(m_2,\pm)$ ($m_1>m_2$). This occurs at the drift velocity 

\begin{equation}
v_{\rm dr}^{(m_1,-),(m_2,\pm)}\approx \frac
{\omega_{p2}(G_{m_1})+\omega_{p2}(G_{m_2})} 
{G_{m_1}\mp G_{m_2}},
\end{equation}
and is accompanied by an enhancement of the amplification of waves,
due to a resonant interaction of different plasma modes. In Figure
\ref{MG_drift2} one sees this effect at $U\approx 5.0$ [the intersection of
modes $(3,-)$ and  $(1,+)$]. A small resonance feature can be also seen
at the intersection of modes $(3,-)$ and $(1,-)$ at $U\approx 2.6$ in the
low-frequency range. 

In vacuum devices one can easily  achieve the drift velocity sufficient
for the amplification of electromagnetic waves. In solid-state structures
``{\it metal} grating - 2DES'' the discussed values of the threshold velocity
are rather large. In order to make realistic estimations of achievable 
drift velocities in semiconductor heterostructures with the 2D electron gas we
refer to the paper of Wirner {\it et al.}, \cite{Wirner} in which the
dependence of the average drift velocity of 2D electrons as a function of the
applied electric field has been experimentally investigated. The authors
studied a GaAs/Al$_x$Ga$_{1-x}$As heterostructure with the density of
2D electrons 
of $n_2=6\times 10^{10}$ cm$^{-2}$ and the (low-field) mobility of
$\mu_2=8\times 10^5$ cm$^2$/Vs. The corresponding Fermi velocity of 2D
electrons is $v_{\rm F2}= 1.06\times 10^7$ cm/s. The measured drift
velocity of 2D electrons increases linearly with the applied electric field up
to $E_0\approx 50$ V/cm, sublinearly at larger fields and then 
saturates at $v_{\rm dr}\approx 1.8\times 10^7$ cm/s when the field is
increased up to $E_0\approx 150$ V/cm. Based on the 
results of this experiment we will assume that the really achievable
experimental values of the ratio $U=v_{\rm dr}/v_{\rm F2}$ in
GaAs/Al$_x$Ga$_{1-x}$As heterostructures with the low-density high-mobility 2D
electron gas are restricted by the value of $U\approx 1.8$.
 
As seen from the above examples the threshold velocity is still well above
the desired limit $U\approx 1.8$. It can be reduced, as compared to the
numerical 
examples discussed above, by using  smaller 2D electron gas density $n_2$,
smaller grating period $a$, and  larger 2D plasmon harmonics $m$. One of the
problems in using the higher  $m$ in the structures ``{\it metal}
grating - 2DES'' is a small amplitude of 2D plasmon resonances with $m>1$ 
 and their rapid decrease with increasing $m$, see Figure \ref{MG_nodrift}. As
we saw however in  Section \ref{nodriftQW}, the use of the {\it quantum
wire} grating allows one to increase the amplitudes of the higher 2D plasmon
resonances by an order of magnitude (compare Figures \ref{QW_nodrift} and
\ref{MG_nodrift}c), at the cost of the resonant interaction of 2D plasmons in
the 2DES and the plasmons in wires. Using this effect, along with other
methods discussed above, one can reduce the threshold velocity down to
experimentally achiavable values. An amplification of electromagnetic waves
in the structure ``{\it quantum wire} grating - 2DES'' is considered in the
next Section.

\subsection{Quantum-wire grating}
\label{driftQW}

In order to make a realistic estimation of the transmission coefficient of
electromagnetic waves 
in the structure ``quantum wire grating - 2DES'' we do this for a hypothetic
sample with parameters taken from published experimental papers. We assume
that our sample is a GaAs/Al$_x$Ga$_{1-x}$As heterostructure ($m_1=m_2=0.067$)
with the density of 2D electrons in the 2DES of $n_2=6\times 10^{10}$
cm$^{-2}$ (taken from Ref.\ \onlinecite{Wirner}). The low-field mobility in
Ref.\ \onlinecite{Wirner} was about $\mu_2\approx 8\times
10^5$ cm$^2$/Vs which corresponds to $\gamma_2\approx 3.25\times
10^{10}$  s$^{-1}$. In the high-field regime ($E_0\approx 150$ V/cm) the
mobility was by a factor of $\sim 4$ smaller, due to a heating of 2D
electrons by a strong dc current. The dependence of the mobility on the dc
current (or on the drift velocity) could be included into the theory through a
phenomenological dependence of the 
momentum relaxation rate $\gamma_2(T_e)$ on the electron temperature. For our
estimations we use, for simplicity,  the drift velocity {\it independent} value $\gamma_2=1.3\times
10^{11}$  s$^{-1}$, which corresponds to the mobility $\mu_2\approx 2\times
10^5$ cm$^2$/Vs (roughly, this equals to the ratio
$v_{\rm dr}/E_0$ at $E_0\approx 150$ V/cm in Ref.\ \onlinecite{Wirner}). Thus we assume the worst value of the momentum
relaxation rate and take effectively into account the
heating of 2D electrons by the strong dc current. For the grating
 we assume the same momentum relaxation rate
of electrons, $\gamma_1=\gamma_2$. 

Choosing the geometrical
parameters of the structure we have assumed that modern experimental
technique allows one to create periodic microstructures with lateral
dimensions of order of $0.1$ $\mu$m, see e.g. Refs.\
\onlinecite{Weiss91,Dahl95}. The width of the grating
strips is therefore taken to be $W=0.1$ $\mu$m, while the period $a=0.175$
$\mu$m is chosen in accordance with the rule (\ref{bestconditions}) for
$m=3$. The distance between the 2DES and the quantum-wire grating is assumed
to be $D=20$ nm. 

Figures \ref{qw1drift} and \ref{qw2drift} demonstrate the calculated
transmission coefficient of the structure ``quantum wire grating -- 
2DES'', near the intersection point of
the unstable 2DES plasma mode $\Omega_{12}(3,-)$ and the grating plasmon. Two
different values of the electron density in the grating, $n_1=n_2=6\times 10^{10}$ cm$^{-2}$ (Figure \ref{qw1drift}) and
$n_1=2n_2=1.2\times 10^{11}$ cm$^{-2}$ (Figure \ref{qw2drift}), are used. 
Three important features
seen on Figures \ref{qw1drift} and \ref{qw2drift} should be
mentioned. First, the resonant amplification of
electromagnetic waves occurs at the drift velocities well below the
experimentally achievable limit $U\approx 1.8$. The value of $U\approx 1.4$
(Figure \ref{qw1drift}) corresponds to the drift velocity $v_{\rm dr}\approx
1.4\times 10^7$ cm/s and the dc current density $j_0\approx 0.13$ A/cm
in our example; in Ref.\ \onlinecite{Wirner} this velocity has been achieved
at $E_0\approx 50$ V/cm. Second, the operating frequency of the amplifier lies
in the vicinity of the intersection point of the grating plasmon
$\tilde\omega_{p1}$ and the unstable mode $\Omega_{12}(m,-)$, at 

\begin{equation}
v_{\rm dr}\approx \frac
{\tilde\omega_{p1}+\omega_{p2}(G_m)\sqrt{1-\Delta_m/2}} 
{G_m}, {\ \ \ } \omega\approx \tilde\omega_{p1}.
\end{equation}
It is varied by the dc current (the drift velocity) within about 10 \% with
respect to $\omega\approx \tilde\omega_{p1}$ if the physical and geometrical
parameters of the structure are kept constant (for instance, from $\approx
0.73$ to $\approx 0.8$ THz when $U$ changes from 1.36 to 1.4 at Figure
\ref{qw1drift}, or from 
$\approx 1.03$ to $\approx 1.13$ THz when $U$ changes from 1.52 to 1.58 at
Figure \ref{qw2drift}). The operating frequency can be also varied by changing
the frequency $\tilde\omega_{p1}$ (compare Figures \ref{qw1drift} and
\ref{qw2drift}), in quantum-wire structures tunable, e.g., by a gate
voltage. Third, the absolute value of the amplification of 
waves near the resonances can be as great as several tens of percents, which
is due exclusively to the resonant interaction of the 2D plasmons with the 
grating plasmon in the {\it quantum wire} grating. For a comparison, in Figure
\ref{mg3drift} we show  
the transmission coefficient of the structure ``{\it metal} grating --
2DES'' for the same parameters of the 2DES, the same geometrical
parameters, and in the same frequency and drift velocity intervals
as in Figure \ref{qw2drift}. A weak resonant feature which intersects the plot
along the diagonal is the unstable mode $\Omega_{12}(3,-)$. As seen from
Figure \ref{mg3drift}, the amplification of waves in the metal grating structure
 is {\it several orders of magnitude} smaller than in
the structure with the quantum wire grating (note the very large
difference in the vertical axis scales in Figures \ref{qw2drift} and
\ref{mg3drift}). 

Thus an amplification of FIR
radiation  in structures with the quantum wire grating can be obtained at realistic, experimentally achievable parameters.

\section{Emission of waves}
\label{emission}

In previous Sections we have discussed an amplification of electromagnetic
waves in the structure influenced by {\it both} the incident electromagnetic
wave {\it and } the strong dc current (stimulated radiation). If the incident
wave is absent, but the sample experiences a current flow, the system emits
electromagnetic waves due to a disturbance of the
thermal equilibrium  (spontaneous radiation). This
situation has been realized 
in experimental papers published so far.
\cite{Tsui80,Gornik80,Hopfel82PRL,Hopfel82,Okisu,Hirakawa95} In order to
describe the emission 
spectrum using the formalism developed above one should include into the
consideration the equilibrium black-body radiation 
around the system, and take into account that the sample has a higher
temperature ($T_e$) than the environment ($T_0$). Taking into account that the
sample reflects and transmits the incident black-body radiation with the
intensity $I_{bb}(\omega,T_0)$, and emits the radiation with the
intensity $A(\omega,v_{\rm dr})I_{bb}(\omega,T_e)$, \cite{LLstat} we obtain
that the 
emitted radiation registered by an external device (filter) in the frequency interval between $\omega$ and $\omega+d\omega$ is given by \cite{note4}  

\begin{equation}
E(\omega,v_{\rm dr})=A(\omega,v_{\rm
dr})[I_{bb}(\omega,T_e)-I_{bb}(\omega,T_0)]. 
\label{emissi}
\end{equation}
Here $A(\omega,v_{\rm dr})$ is the absorption coefficient of the structure
calculated in 
Section \ref{nodrift}, Eq. (\ref{absorption}), and 

\begin{equation}
I_{bb}(\omega,T_0)=\frac{\hbar \omega^3d\omega}{4\pi
c^2(e^{\hbar \omega/T_0}-1)},
\label{bb}
\end{equation}
is the intensity of the black-body radiation in the interval $(\omega,\omega+d\omega)$. 

Figure \ref{emissionspectrum} demonstrates the
absorption (thin curves) and emission (thick curves) spectra of the structure
``metal grating -- 2DES'' under the conditions of the experiment of Ref.\
\onlinecite{Hirakawa95} ($n_2=5.4\times 10^{11}$ cm$^{-2}$, $a=2$ $\mu$m and
$a=3$ $\mu$m for two different samples, $W/a=0.6$ in both cases, $D=62$
nm, and the filter linewidth $\Delta f=85$ GHz). Plotting the Figure we have assumed a step-like electron density 
profile in the grating, as the more relevant one in the case of the metal grating with wide strips, as well as the scattering rate 
$\gamma_2=5\times 10^{11}$ s$^{-1}$, 
the environment temperature $T_0=4.2$ K, and $T_e=100$ K ($T_e=50$ K) for the
 sample with $a=2$ $\mu$m ($a=3$ $\mu$m). The drift velocity was
assumed to be vanishing.  The whole
behaviour of the emission spectra qualitatively agrees with the measured
ones, the position of peaks in Figure \ref{emissionspectrum} is in a good
quantatitive 
agreement with those measured in Ref.\ \onlinecite{Hirakawa95}, see Table \ref{table}. 
 
It should be noted that the problem of the emission of light from the grating
coupled 2DES with a flowing current has been considered in Ref.\
\onlinecite{Kempa93}. It has been solved in a complete analogy with the
transmission problem. Such formulation of the emission problem is however not
well defined and cannot give the emission spectrum (\ref{emissi}) (the fact
that the emission of light from the system is due to the black-body radiation
of a sample heated by the dc current has been ignored in Ref.\
\onlinecite{Kempa93}). Indeed, when the transmission of light is calculated,
one gets a set of equations for Fourier components of the total electric field
(Section \ref{formalism} or Ref.\ \onlinecite{Zheng})

\begin{equation}
\sum_{{\bf G}^\prime}\hat M_{{\bf G,G}^\prime} E_{{\bf G}^\prime}^{\rm tot}=E_{\bf G}^{\rm ext}, 
\label{matrix}
\end{equation}

\noindent
where $\hat M_{{\bf G,G}^\prime}$ is an infinite matrix over  reciprocal
lattice vectors. The spectrum of eigen  modes is determined by the equation
$\det \hat M=0$, the total self-consistent electric field can be found from
Eq. (\ref{matrix}), if the  matrix $\hat M$ is inverted. \cite{Zheng} The {\it
transmission} problem is thus well defined. 

In Ref.\ \onlinecite{Kempa93} the authors have derived a similar equation for
the {\it emission} problem, when $v_{\rm dr}\neq 0$ and ${\bf E}^{\rm
ext}=0$. In this case however, the right-hand side of Eq. (\ref{matrix})
vanishes, and the drift velocity $v_{\rm dr}$ enters only the matrix $\hat
M$. The spectrum of eigen modes of the system (including unstable ones) can be
calculated from the equation $\det \hat M=0$, but the induced radiated field
$E_{\bf G=0}^{\rm ind}$ cannot be found in this fashion, as the right-hand
side of Eq. (\ref{matrix}) is zero. Instead of the intensity of the emitted
waves, Kempa {\it et al.} \cite{Kempa93} calculated the ratio of the
macroscopic (radiated) field $E_{\bf G=0}^{\rm ind}$ to the microscopic
(nonpropagating) field $E_{\bf G\neq 0}^{\rm ind}$. This ratio characterizes
the grating as a coupler of the plasmon field to the propagating
electromagnetic radiation, but is not an appropriate characteristics of the
emission process, as it does not vanish, for instance, at $v_{\rm dr}=0$.

\section{summary}
\label{discussion}

We have developed a general analytic theory of the transmission,
reflection, absorption and emission of electromagnetic waves in the structure
``2DES (with and without the flowing dc current) -- grating coupler''. We have
demonstrated that an amplification, and hence a generation, of FIR radiation
(the lasing effect) can be obtained in semiconductor microstructures with
realistic experimental parameters. Summarizing our results we formulate the
requirements which should be met in order to create a successfully working
amplifier of FIR radiation based on the grating coupled 2DES with a flowing
current. 

\begin{enumerate}
\item The density of electrons in the 2DES should be small. This requirement
seems to be paradoxical, as the intensity of the transmission resonance at the
vanishing drift velocity $v_{\rm dr}=0$ becomes very small when $n_2$
decreases. Nevertheless, at a large drift velocity the intensity of resonances
becomes sufficiently large (Figure \ref{MG_drift2}), especially in the structures with a quantum wire
grating (Figures \ref{qw1drift}, \ref{qw2drift}), but the threshold velocity of
amplification decreases with $n_2$, Eqs. (\ref{threshveloc}), (\ref{1stcontrib}). Note
that this requirement has not been satisfied in previous emission experiments
(for instance, in Ref.\ \onlinecite{Hirakawa95} the 2D electron
density was by an order of magnitude larger than the value that we have used
in Figures \ref{qw1drift}, \ref{qw2drift}). Note also that in vacuum devices
the plasma frequency in the electron beam used to be very small ($G_1v_{\rm
dr}\gg \omega_p$). 

\item The mobility of 2D electrons should be sufficiently large, however this
requirement is not so crucial as others. As we have seen in Figures
\ref{qw1drift} and \ref{qw2drift}, a considerable amplification of light could
be achieved at a moderate mobility of $\mu_2\approx 2\times 10^5$ cm$^2$/Vs.

\item The most important requirement imposed on the grating is that it must be
a {\it quantum wire} grating, but not a commonly employed
\cite{Tsui80,Gornik80,Hopfel82PRL,Hopfel82,Okisu,Hirakawa95} metal one. It is
the resonant interaction of plasma modes in the 2DES and in the grating that
allows one significantly to increase the amplification of light when using the
higher 2D plasmon harmonics, and hence to reduce the threshold velocity of
amplification. 

\item The grating period should be as small as possible. This condition
seems to be in some contradiction with another one, $a\gg D$, which follows
from the requirement that the interaction of plasma modes in the 2DES and in
the grating [described by the exponent $\exp(-G_mD)$, Eq. (\ref{Delta})] 
should be sufficiently large (for instance, in Ref.\ \onlinecite{Hirakawa95}
the grating period was  by an order of magnitude larger than
the one we have used in Figures \ref{qw1drift}, \ref{qw2drift}). Nevertheless,
the inequality $D\ll a$ (more accurately, $D\ll W$, see below) should be
considered as a condition for $D$, while the period should be taken to be
small. As seen from Eqs. (\ref{1stcontrib}), (\ref{2ndcontrib}), this leads to
a reduction of the threshold velocity. 

\item The width of the quantum wires {\it must satisfy} the conditions
(\ref{bestconditions}) (or, $Wm/a=1,2,3,...$, if the step-like profile seems to
be more appropriate for a description of a particular system \cite{note5}). As
seen from Figure \ref{threshvel}, a correct choice of the ratio $W/a$ {\it is
of particular importance}, especially when the higher 2D plasmon harmonics are
used.  

\item The distance $D$ between the 2DES and the grating should meet the
condition $D\ll W$. This requirement follows from maximizing 
the exponent $\exp(-G_mD)$, Eq. (\ref{Delta}), if to take into account that
the ratio $Wm/a$ is already fixed by the conditions (\ref{bestconditions}). 

\item The operating frequency of amplifiers can be varied by the dc electric
current flowing in the 2DES and/or by changing the electron density $n_1$ (more
generally, the resonance frequency $\tilde\omega_{p1}$) in the quantum wire
grating. 

\end{enumerate}

We hope that these recommendations will finally help to 
create tunable FIR sourses and lasers, based on electron semiconductor
microstructures of low dimensionality.

\acknowledgments

I thank N. Savostianova for many useful discussions and W. L. Schaich for
sending me the preprint of his paper \cite{schaich} before publication.

\appendix

\section{solving the integral equation (\ref{INTEQ})}
\label{derivation}

A general scheme of solving the integral equation (\ref{INTEQ}) consists in
the following. Let $O_n(x)$ is a set of orthogonal polinomials with respect to
the weight function $\vartheta(x)$ which satisfy the condition

\begin{equation}
\int\frac{dx}W\vartheta(x)O_n(x)O_m(x)=\delta_{mn}
\end{equation}
[for the profile (\ref{elliptic}) $O_n(x)$ are the Chebyshev
polinomials]. Substituting an expansion 

\begin{equation}
E_x^{\rm tot}(x)=\sum_nC_nO_n(x)
\end{equation} 
into Eq. (\ref{INTEQ}), multiplying by $\vartheta(x)O_m(x)$ and
integrating over $dx$ we get an infinite set of equations

\begin{equation}
\sum_n(\delta_{mn}+L_{mn})C_n=E_0W({\bf 0},\omega)\delta_{m0},
\end{equation}
which can be solved approximately truncating the matrix $\delta_{mn}+L_{mn}$
to a finite size $N\times N$. Here  

\begin{equation}
L_{mn}=\frac{2\pi if \langle \sigma_{1D}(\omega)\rangle}{\omega\epsilon} 
\sum_{\bf G} W({\bf G},\omega) \kappa_G \beta_m({\bf G})\beta^\star_n({\bf G}),
\end{equation}   
$\beta_m({\bf G})=\langle \vartheta(x)O_m(x) e^{i{\bf G}\cdot{\bf r}}
\rangle$, and the star means the complex conjugate. The approximation accepted
in the main body of the paper corresponds to 
$N=1$. Physically, truncating the matrix to the $1\times 1$ size we neglect
effects related to an excitation of quadrupole and higher eigen plasma modes
\cite{aleiner} in the grating strips.  

\section{resonance frequency in an array of wires}
\label{wires}

Equation (\ref{1Darray}) gives an expression for a dipole excitation resonance
frequency in an array of wires as a functional of the equilibrium electron
density $\vartheta(x)$. It can be presented in different forms which clarify
the role of the inter-wire interaction and the relation to the generalized
Kohn theorem. \cite{brey} 

\subsection{Inter-wire interaction}

In view of Eq. (\ref{singlewire}) one can write

\begin{equation}
\frac{\tilde\omega_{p1}^2}{\omega_{p1}^2}=\frac{\pi fW}8\sum_{G_m\neq 0}
|G_m| \alpha(G_m).
\end{equation} 
For a semielliptic density profile (\ref{elliptic}) the form-factor $\alpha(G)$
is given by Eq. (\ref{ffelliptic}). Using the transformation 

\begin{equation}
\sum_{G_m} F(G_m) = \int{a dq \over{2\pi}} F(q)
\sum_k e^{iqa_k}, 
\label{Poistrans}
\end{equation}
where $a_k=ak$, we get

\begin{equation}
\frac{\tilde\omega_{p1}^2}{\omega_{p1}^2}=\int_{-\infty}^\infty \frac{dq}{|q|}
J_1^2(qW/2)\sum_{k=-\infty}^\infty e^{iqak}.
\end{equation}
The term of the sum with $k=0$ is independent of the grating period $a$ and
gives, after the integration over $dq$, unity. It describes the contribution
of a single wire. The corrections due to the inter-wire interaction are then
written as

\begin{equation}
\frac{\tilde\omega_{p1}^2}{\omega_{p1}^2}=1+2\sum_{k=1}^\infty\int_0^\infty
\frac{dx}{x}J_1^2(fx/2k)(e^{ix}+e^{-ix}).
\label{2int}
\end{equation}
In the first (second) integral in Eq. (\ref{2int}) we rotate the integration
path by an angle $+\pi/2$ ($-\pi/2$) to the upper (lower) complex
half-plane. The function $J_1$ is transformed to $I_1$, and we have

\begin{equation}
\frac{\tilde\omega_{p1}^2}{\omega_{p1}^2}=1-4\sum_{k=1}^\infty\int_0^\infty
\frac{dx}{x}e^{-x}I_1^2(fx/2k).
\label{expan}
\end{equation}
The expansion (\ref{1Dresfreq}) can now be easily obtained from
Eq. (\ref{expan}) at $f\ll 1$.

\subsection{Relation to the generalized Kohn theorem}

In many publications concerning the electromagnetic response of a confined
system of electrons like quantum wires or quantum dots, the resonance
frequencies are discussed in terms of not an equilibrium electron density, see
Eq. (\ref{1Darray}), but of a confining potential. One
can state a relation between these two approaches.

Let an array  of wires is formed by an external confining potential (potential energy)

\begin{equation}
V_{\rm ext}(x)=\sum_{k} v_{\rm ext}(x-a_k).
\label{ext}
\end{equation}

\noindent
The total self-consistent potential $V_{\rm tot}(x)=V_{\rm ext}(x)+V_{\rm
ind}(x)$ is given by the sum of external and induced potentials, where the
induced potential $V_{\rm ind}(x)$ relates to the density (\ref{density}) by
the Poisson equation 

\begin{equation}
\Delta V_{\rm ind}(x,z)=-\frac{4\pi e^2 N_1(x)}{\epsilon} \delta(z).
\label{PoissonEq}
\end{equation}
Equation (\ref{1Darray}) can be written as

\begin{equation}
\tilde\omega_{p1}^2 = \frac{2\pi e^2}{m_1\epsilon Wn_1} \sum_{m\neq 0}
|G_m| N_{1,G_m}\int_{\rm cell}dxn_1(x)e^{iG_mx},
\label{1Darray0}
\end{equation} 
where the integral is taken over an elementary cell. Due to
Eq. (\ref{PoissonEq}) the Fourier component of the density $N_{1,G}$ is
related to the Fourier component of the induced potential
$V_{{\rm ind},G}=2\pi e^2N_{1,G}/\epsilon|G|$, so that we can write
Eq. (\ref{1Darray0}) for a single wire in the form

\begin{equation}
\omega_{p1}^2 = -\frac{1}{m_1n_1W} \int dxn_1(x)\Delta_2V_{\rm ind}(x),
\label{1Darray1}
\end{equation} 
where $\Delta_2$ is the two-dimensional Laplacian and the integral is expanded
onto the whole axis. Replacing $V_{\rm ind}$ by the difference $V_{\rm
tot}-V_{\rm ext}$, one sees that the contribution due to the total
self-consistent potential vanishes, as $V_{\rm tot}$ is constant in
points where electrons are. Thus, we have

\begin{equation}
\omega_{p1}^2 = \frac{1}{m_1n_1W} \int dxn_1(x)\Delta_2V_{\rm ext}(x).
\label{1Darray2}
\end{equation} 
In a parabolic confining potential $V_{\rm ext}(x)=Kx^2/2$, and 
Eq. (\ref{1Darray2}) reproduces the statement of the generalized Kohn theorem,
\cite{brey} 

\begin{equation}
\omega_{p1}^2 = K/m_1.
\end{equation}



\begin{figure}
\caption{The geometry of the considered structure. The system in infinite in
$y$ direction. 2D electrons are moving in the $x$ direction perpendicular to
the grating strips with the drift velocity $v_{\rm dr}$. A transmission
spectroscopy experiment corresponds to $v_{\rm dr}=0$, $I_0\neq 0$, an
emission spectroscopy experiment corresponds to $v_{\rm dr}\neq 0$, $I_0=0$,
where $I_0$ is the intensity of the incident wave.}
\label{geometry}
\end{figure}



\begin{figure}
\caption{The transmission, reflection and absorption coefficients of a
quantum-wire grating with $n_1=3\times 10^{11}$ cm$^{-2}$, $m_1=0.067$ (GaAs),
and two different values of the relaxation rate: (a)
$\gamma_1=1.33\times10^{11}$ s$^{-1}$ (corresponds to
$\gamma_1/\Gamma_1=5.0$
), and (b) $\gamma_1=0.53\times10^{10}$ s$^{-1}$
($\gamma_1/\Gamma_1=0.2$
). The ratio $W/a=0.4$.} 
\label{qwgratingonly}  
\end{figure}



\begin{figure}
\caption{The transmission coefficient of the structure ``metal grating --
2DES'' 
at three different values of the ratio $W/a$. Geometrical parameters: $a=1$
$\mu$m, $D=60$ nm. Parameters of the 2D layer: $n_{s2}=3\times 10^{11}$
cm$^{-2}$, $\gamma_2=0.7\times 10^{11}$ s$^{-1}$, $m_2=0.067$ (mobility
$\mu_2=375000$ cm$^2$/Vs). The grating parameters ($n_{s1}=6\times 10^{18}$
cm$^{-2}$, $\gamma_1=1.1\times 10^{14}$ s$^{-1}$, $m_1=1$) correspond to a
typical (Au) grating coupler. The dielectric constant is
$\epsilon=12.8$. Triangles show the calculated positions of the 2D plasmon
harmonics (\protect\ref{2Dplasmons}) for $m=1,\dots,4$. Note the difference of
the vertical axis scales for different plots.}  
\label{MG_nodrift}
\end{figure}



\begin{figure}
\caption{Normalized resonance frequencies $\Omega_{12}(m)/\omega_{p2}(G_1)$ for three different modes $m=1$, 2, and 3 as a
function of the geometrical filling factor 
$f=W/a$. The ratio $D/a=0.08$. Note the difference of the vertical axis scales
for different modes.} 
\label{frequencyfigure}
\end{figure}



\begin{figure}
\caption{(a) Normalized radiative decay of the modes $\Omega_{12}(m)$ and (b)
the resonant transmission coefficient 
$T_{\rm res}(m)$, for three different modes $m=1$, 2,
and 3  as a function of the geometrical filling 
factor $f=W/a$. The ratio $D/a=0.08$, the scattering rate of electrons 
in the 2DES (in Figure b) is $\gamma_2=0.7\times10^{11}$ s$^{-1}$.} 
\label{Tminfigure}
\end{figure}



\begin{figure}
\caption{The transmission coefficient of the structure ``quantum wire grating
-- 2DES'' at three different values of the grating plasmon frequency
$\tilde\omega_{p1}$ at $a=1$ $\mu$m, $D=60$ nm, and $W=0.2$ $\mu$m. Parameters
of the 2D layer and the dielectric constant $\epsilon$ are the same as in
Figure \protect\ref{MG_nodrift}. Parameters of the grating: $\gamma_1=0.7\times
10^{11}$ s$^{-1}$, $m^*_1=0.067$, the electron
density: (a) $n_{s1}=4\times 10^{11}$ cm$^{-2}$, (b) $n_{s1}=2.5\times
10^{11}$ cm$^{-2}$, (c) $n_{s1}=1\times 10^{11}$ cm$^{-2}$. Triangles show the
calculated positions of the 2D plasmon 
harmonics (\protect\ref{2Dplasmons}) for $m=1,\dots,4$, open triangles show
the positions of the grating plasmon.} 
\label{QW_nodrift}
\end{figure}



\begin{figure}
\caption{The absorption coefficient of the structure ``metal grating -- 2DES''
at the frequency interval corresponding to the first ($m=1$) 2D plasmon
harmonic, and at small values of the dimensionless drift velocity $U=v_{\rm
dr}/v_{F2}$. Physical and geometrical parameters of the structure are the same
as in Figure \protect\ref{MG_nodrift}a. The black triangle at the bottom of
the plot shows the position of the $m=1$ 2D plasmon harmonic 
(\protect\ref{2Dplasmons}). The weak mode which intersects the pronounced
resonances at $U\approx 0.8$ and $U\approx 1.5$ is the mode $(3,-)$.}
\label{MG_drift}
\end{figure}



\begin{figure}
\caption{(a) The resonance frequency $\Omega_{12}(m,\pm)$ and (b) the
normalized radiative decay $\Gamma_{12}(m,\pm)/\Gamma_{12}(m)$ at $m=1$ as a
function of the dimensionless drift velocity of the 2D electrons $U=v_{\rm
dr}/v_{F2}$. Physical and geometrical parameters of the structure are the same
as in Figure \protect\ref{MG_drift}.}
\label{rfdec_drift}
\end{figure}



\begin{figure}
\caption{The resonant values of (a) the transmission, (b) the reflection, and
(c) the absorption coefficients for the modes
$\Omega_{12}(m,\pm)$ at $m=1$ as a 
function of the dimensionless drift velocity of the 2D electrons $U=v_{\rm
dr}/v_{F2}$. Physical and geometrical parameters of the structure are the same
as in Figure \protect\ref{MG_drift}.}
\label{TRAres_drift}
\end{figure}



\begin{figure}
\caption{The transmission coefficient of the structure ``metal grating --
2DES'' at the frequency interval corresponding to the first ($m=1$) 2D plasmon
harmonic in a wide range ($0\leq U\leq 8$) of the dimensionless drift velocity
$U=v_{\rm dr}/v_{F2}$. Physical and geometrical parameters of the structure are the same
as in Figure \protect\ref{MG_nodrift}a.}
\label{MG_drift2}
\end{figure}



\begin{figure}
\caption{Threshold velocity of the amplification as a function of the grating
filling factor $f=W/a$ for several lowest mode numbers $m$, and at
$n_{s2}=3\times 10^{11}$ 
cm$^{-2}$, $\gamma_2=0.7\times 10^{11}$ s$^{-1}$, $D=60$ nm. Thin lines show
the first contribution to the threshold velocity (\protect\ref{1stcontrib}).}
\label{threshvel}
\end{figure}



\begin{figure}
\caption{Transmission coefficient of the structure ``quantum wire grating --
2DES'' for $n_1=n_2=6\times 10^{10}$ cm$^{-2}$, $\gamma_1=\gamma_2=1.3\times
10^{11}$ s$^{-1}$, $a=0.175$ $\mu$m, $W=0.1$ $\mu$m, and $D=20$ nm. The
frequency and the drift velocity intervals correspond to an intersection of
the grating plasmon and the $(3,-)$ 2D plasma mode. The black triangle at the
bottom of the plot shows the position of the grating plasmon
(\protect\ref{1Darray}).} 
\label{qw1drift}
\end{figure}



\begin{figure}
\caption{The same as in Figure \protect\ref{qw1drift}, but for 
$n_1=2n_2=1.2\times 10^{11}$ cm$^{-2}$. }
\label{qw2drift}
\end{figure}



\begin{figure}
\caption{The same as in Figure \protect\ref{qw2drift}, but for a metal grating
with $n_{s1}=6\times 10^{18}$
cm$^{-2}$, $\gamma_1=1.1\times 10^{14}$ s$^{-1}$, and $m_1=1$. Note the large
difference in the vertical axis scale in this Figure and in Figure
\protect\ref{qw2drift}.}
\label{mg3drift}
\end{figure}



\begin{figure}
\caption{The absorption (thin curves) and emission (thick curves) spectra of
the structure ``metal grating -- 2DES'' for parameters taken from Ref.\
\protect\onlinecite{Hirakawa95}. The curves are vertically shifted for
clarity, the absorption and emission are plotted in arbitrary units.}
\label{emissionspectrum}
\end{figure}


\begin{table}
\caption{Position of resonance peaks (THz) for two samples with the period
$a=2$ $\mu$m and $a=3$ $\mu$m, measured in Ref.\
\protect\onlinecite{Hirakawa95} and calculated in this work, see Figure
\protect\ref{emissionspectrum}. For other parameters see the text.}
\label{table}
\begin{tabular}{llll}
 & $m=1$ & $m=2$ & $m=3$ \\ \hline
$a=2$ $\mu$m, exp. & 0.69 & 1.3 & --\\
$a=2$ $\mu$m, theor. & 0.68 & 1.24 & --\\ 
$a=3$ $\mu$m, exp. & 0.47 & 0.9 & 1.26\\
$a=3$ $\mu$m, theor. & 0.50 & 1.00 & 1.25\\ 
\end{tabular}
\end{table}


\begin{references}
\bibitem{note0} An assumption on the infinitely thin grating strips implies
that the thickness of strips is small as compared to the skin-layer
thickness. This approximation holds for a quantum wire grating. For a metal
grating this means that the grating strips should be transparent.
\bibitem{note1} When a dc current flows in the system the absorption
coefficient consists in two contributions. The first one in due to the dc
current and is proportional to the real part of the static conductivity of the
2DES, the second contribution is due to the electromagnetic
wave. Eq. (\ref{absorption}) determines only the second contribution to the
total absorption in the system.
\bibitem{note2} In Eq. (\ref{euler}) we have neglected the term describing the
finite compressibility of the 2DES (see Ref.\ \onlinecite{Fetter}). This
approximation holds if $a\gg a_B^\star$, where $a_B^\star$ is the effective
Bohr radius.
\bibitem{note3} Alternatively, the mode $\Omega_{12}(m,-)$ can be classified
at a large drift velocity, $G_mv_{\rm dr}>\omega_{p2}(G_m)$, as the mode
$(-|m|,+)$.
\bibitem{note4} In this estimation we ignore the angular dependence of the
absorption coefficient, assuming that $A(\omega,\theta)=A(\omega,0)$, where
$\theta$ is the angle between the normal to the plane of the 2DES and the
direction of propagation of the black-body radiation. Eqs. (\ref{emissi}) and
(\ref{bb}) are obtained by an integration over $\theta$ of the corresponding
angular dependent expressions.
\bibitem{note5} We believe that the semielliptic density profile is appropriate
for quantum wires, while the step-like profile is more suitable for metal
strips. In a general case, the factor $f^2\alpha(G_m)$ should have a maximum,
see Eq. (\ref{Delta}).
\end{references}
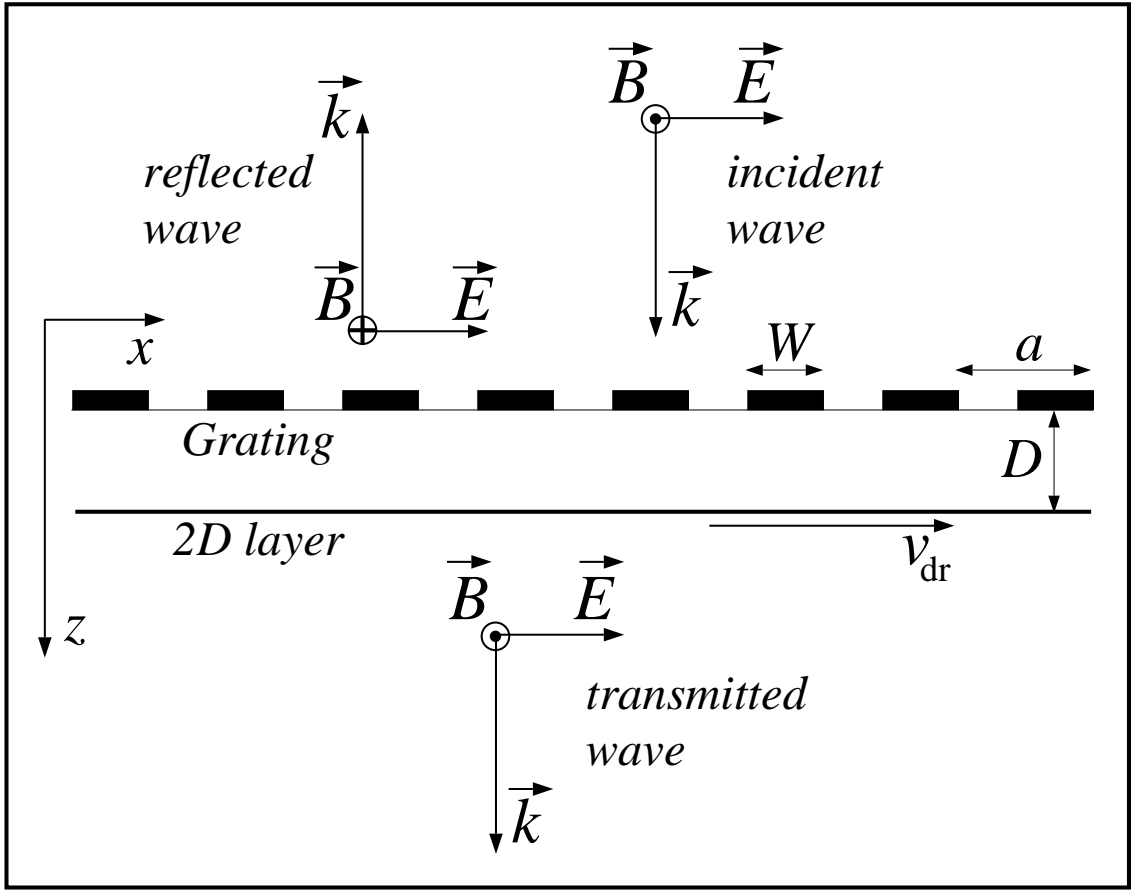
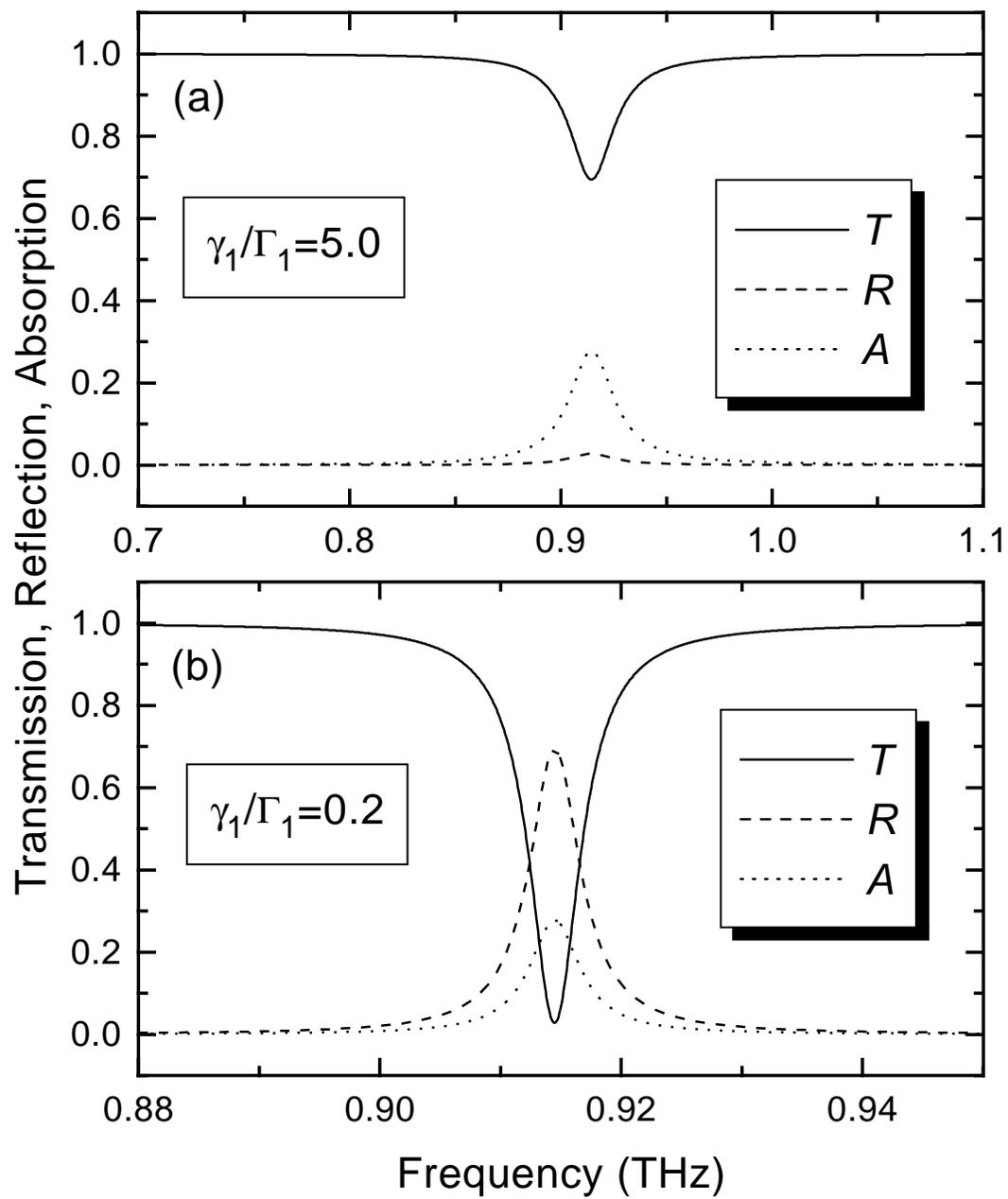
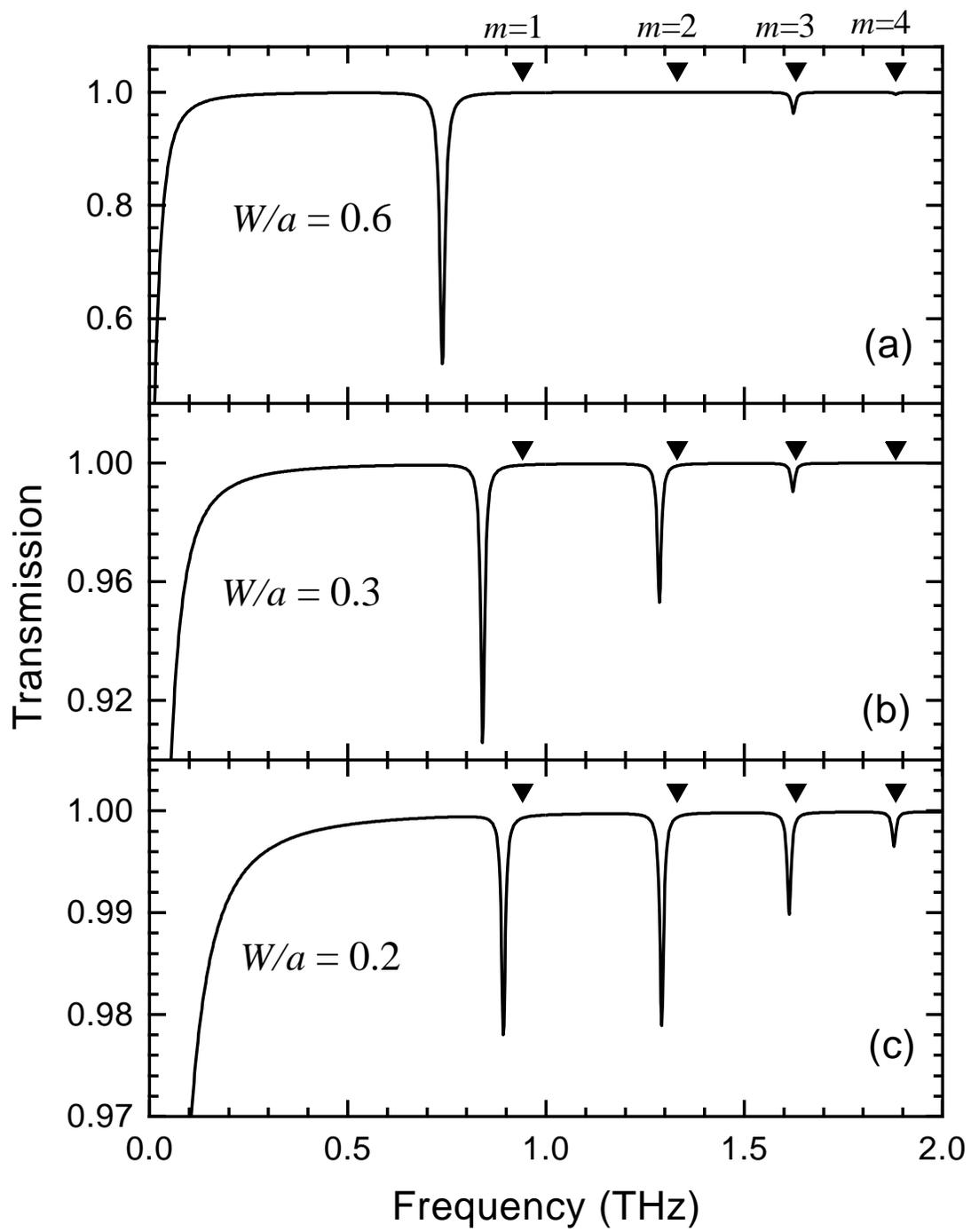
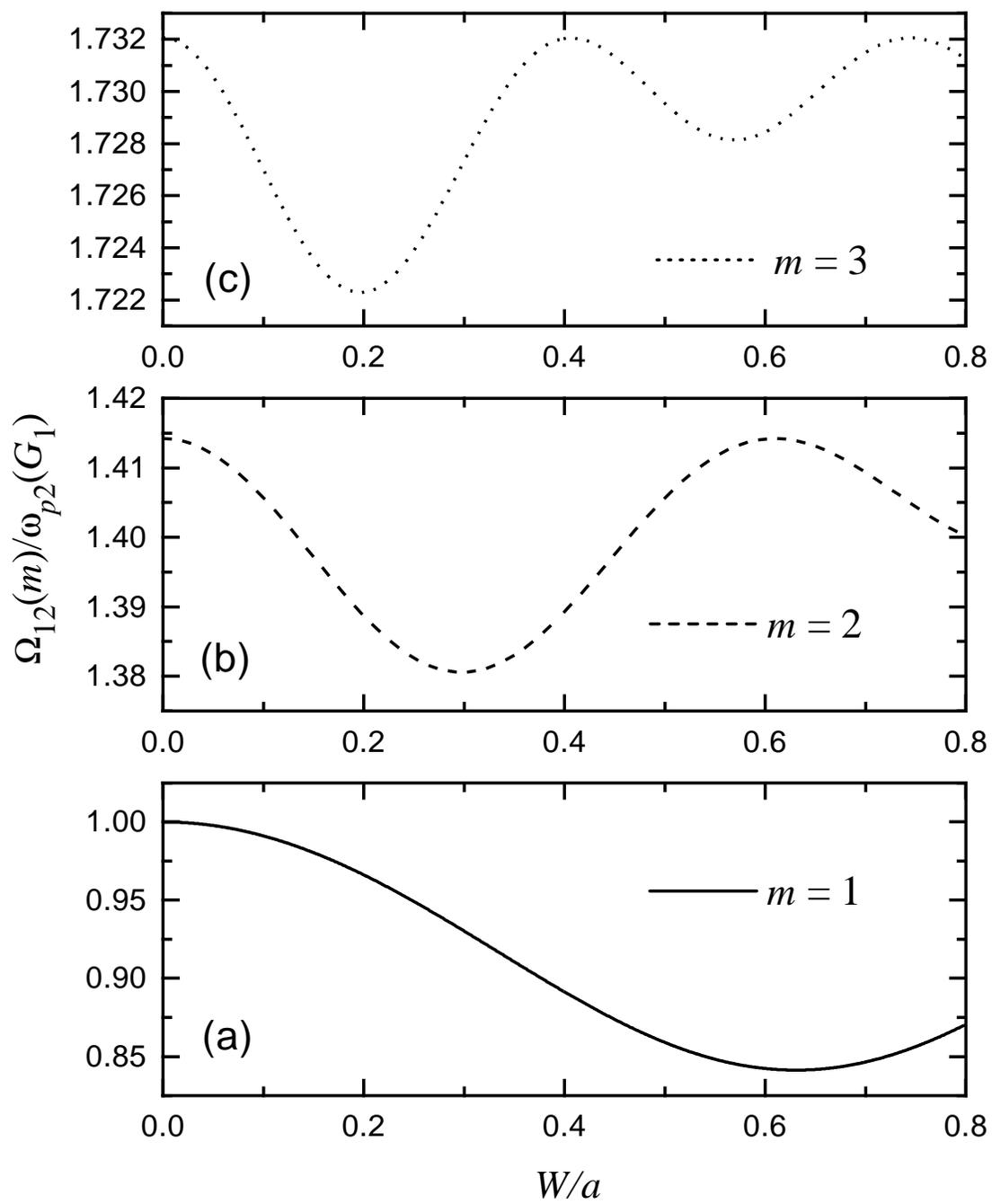
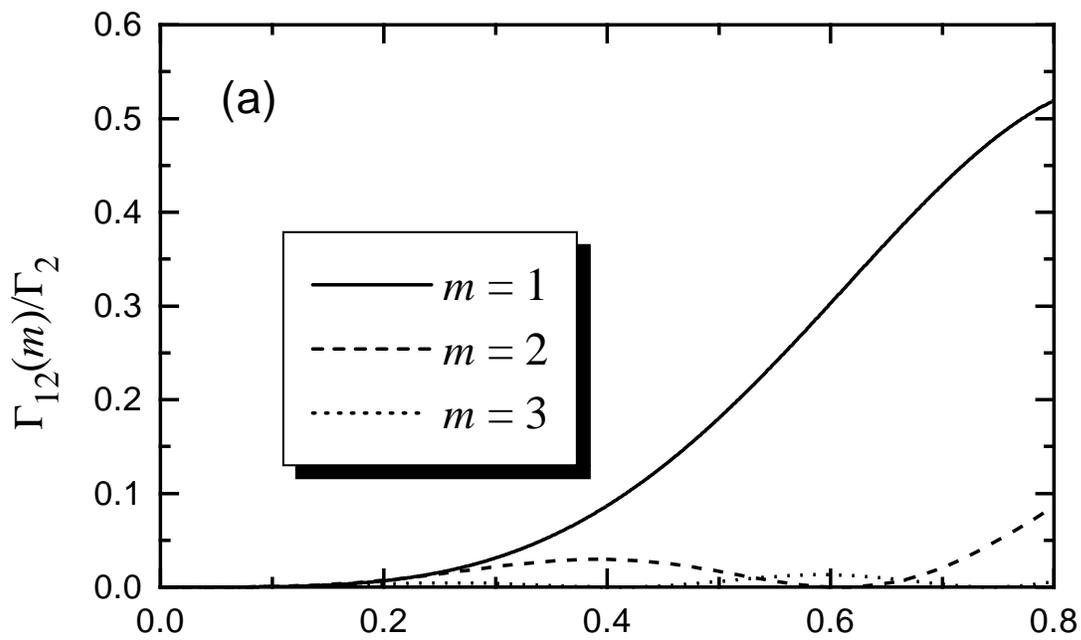
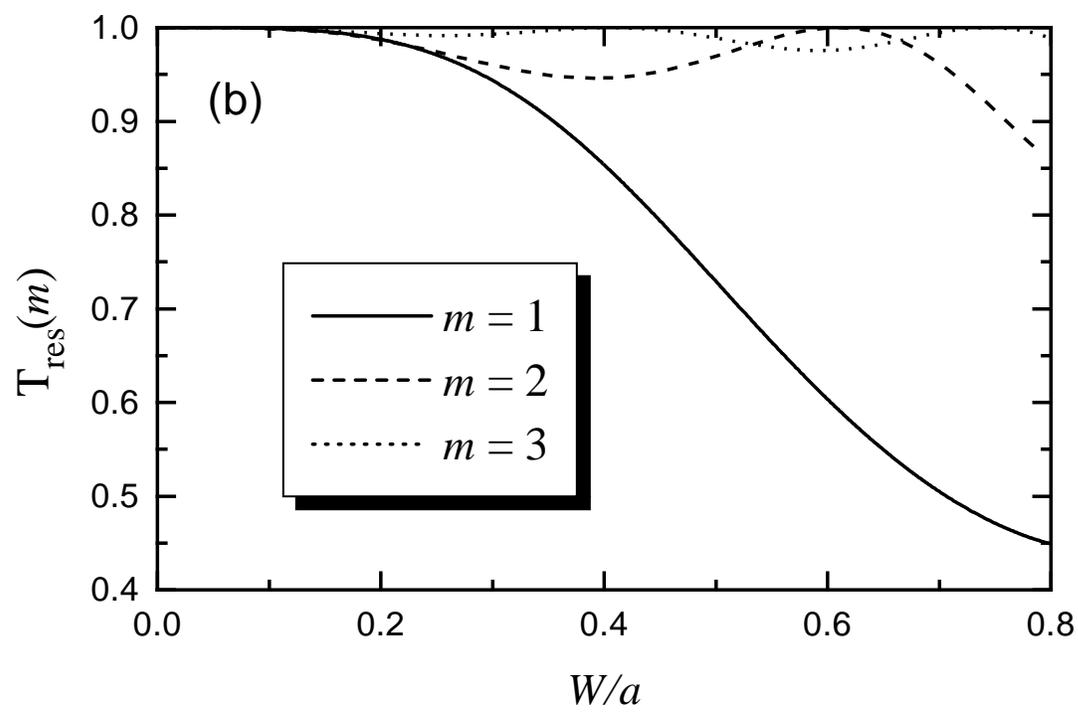
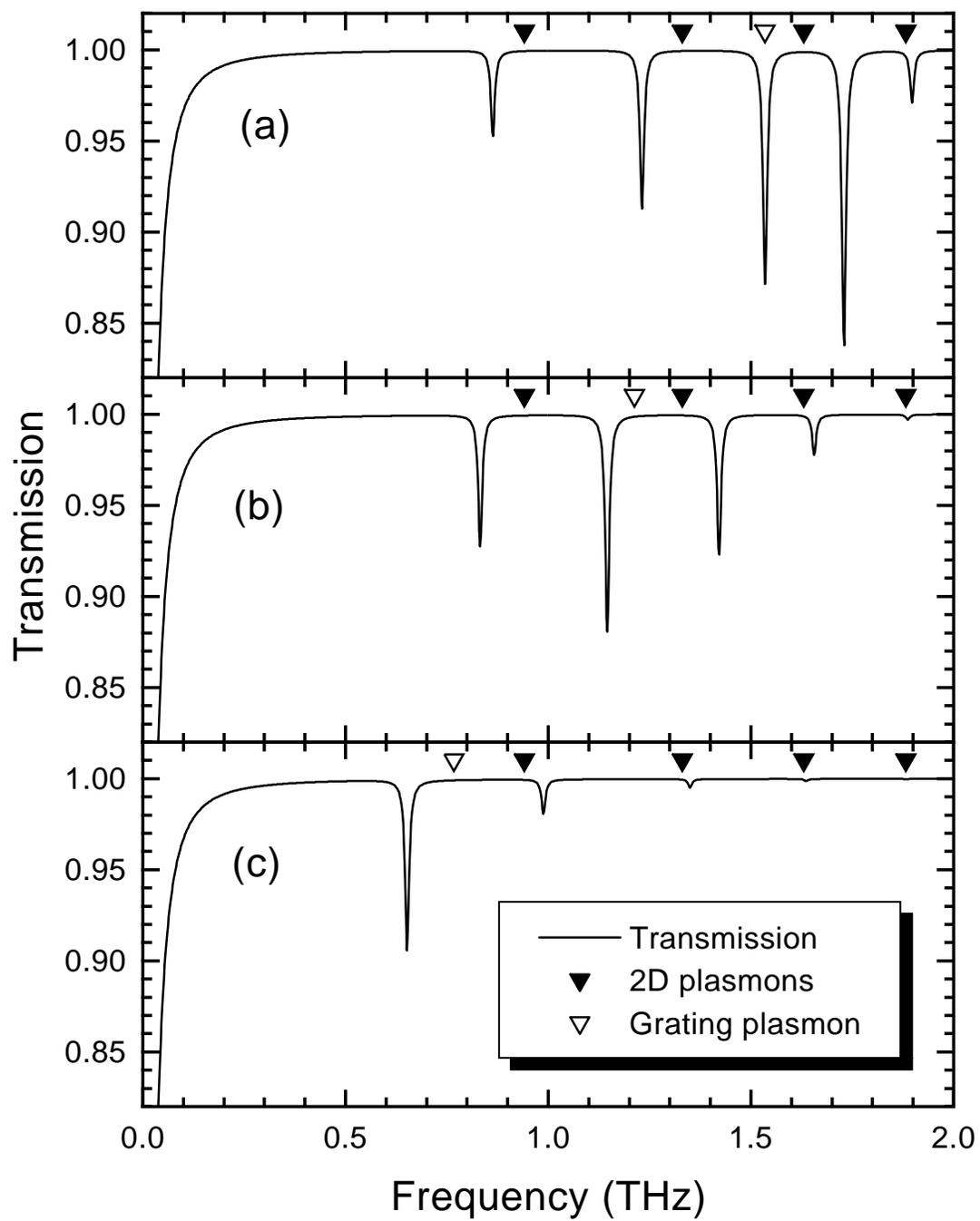
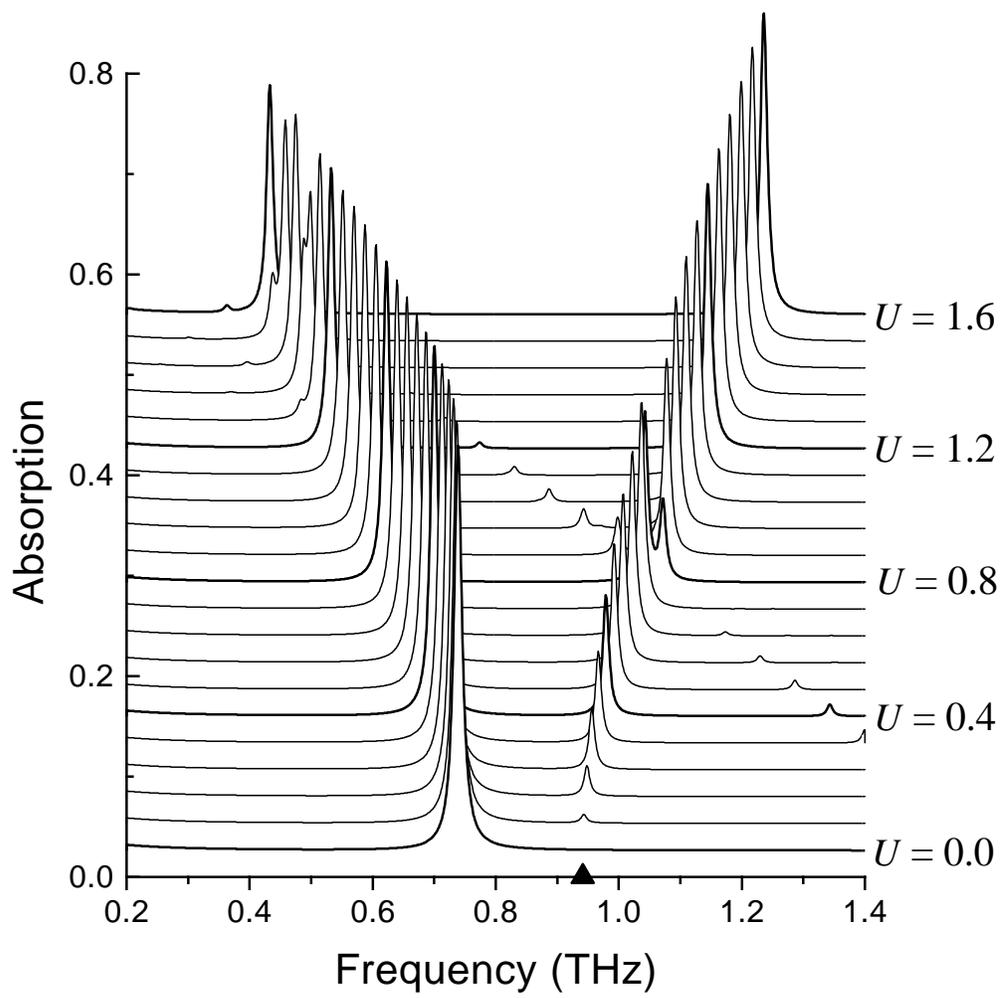
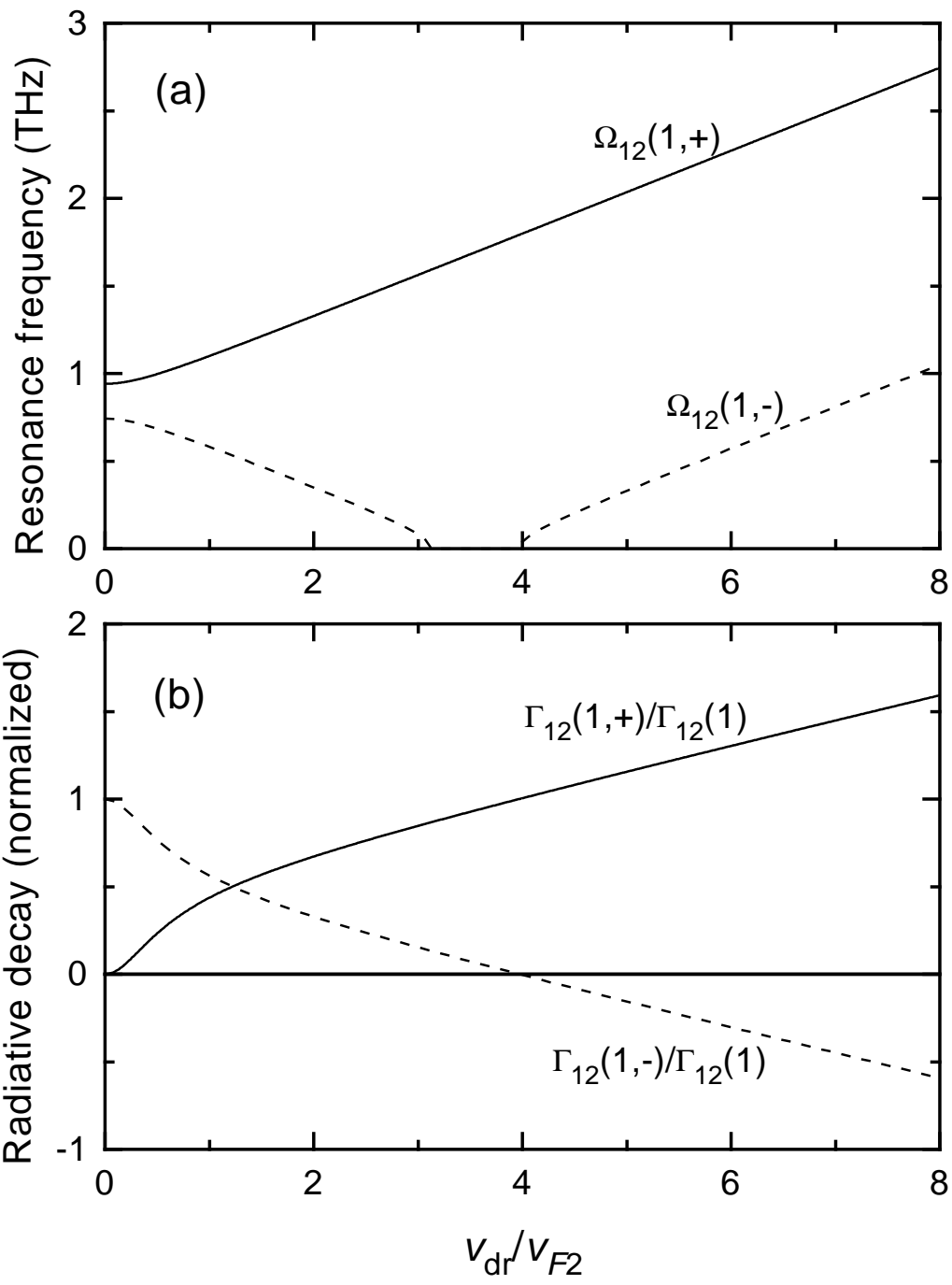
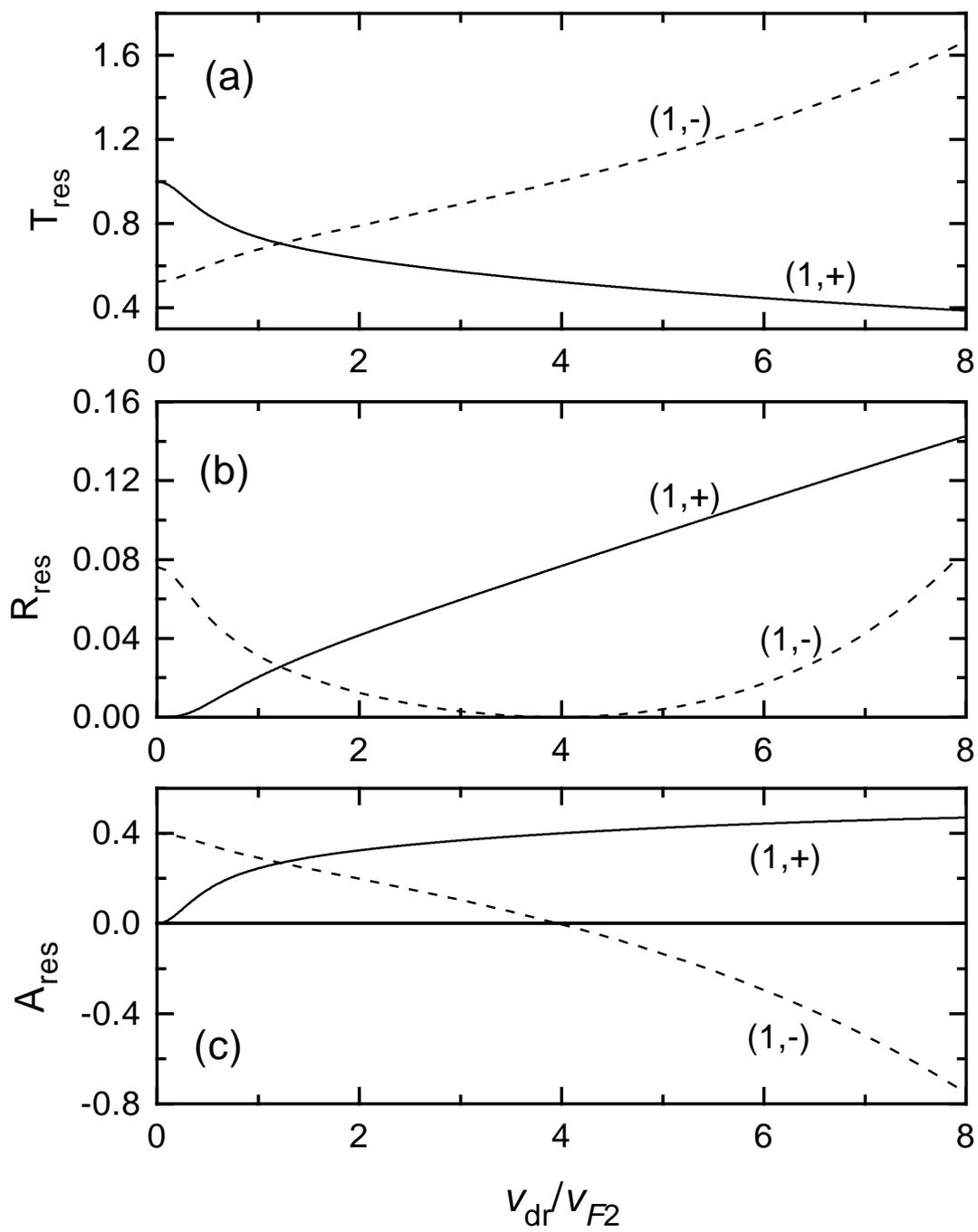
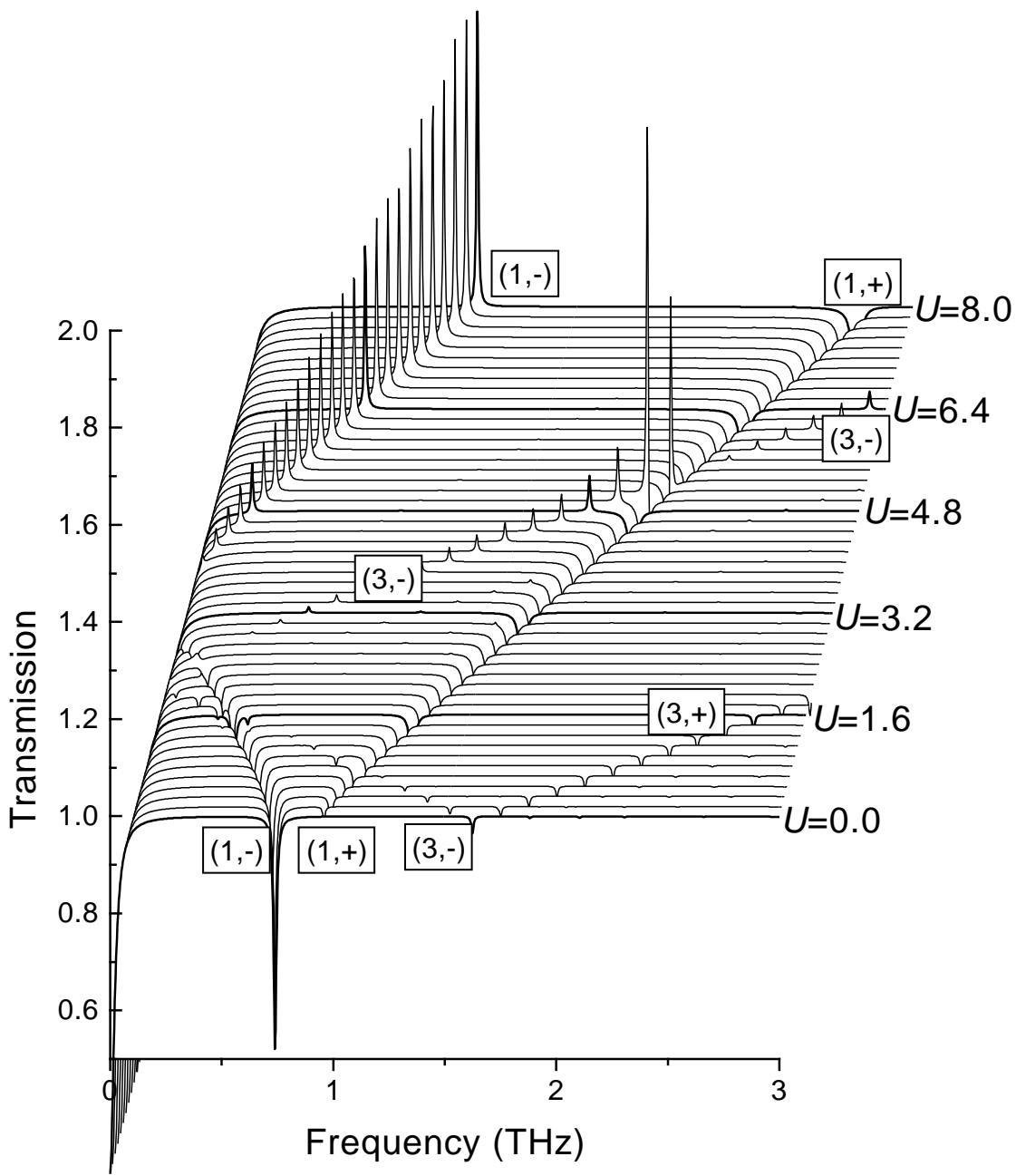
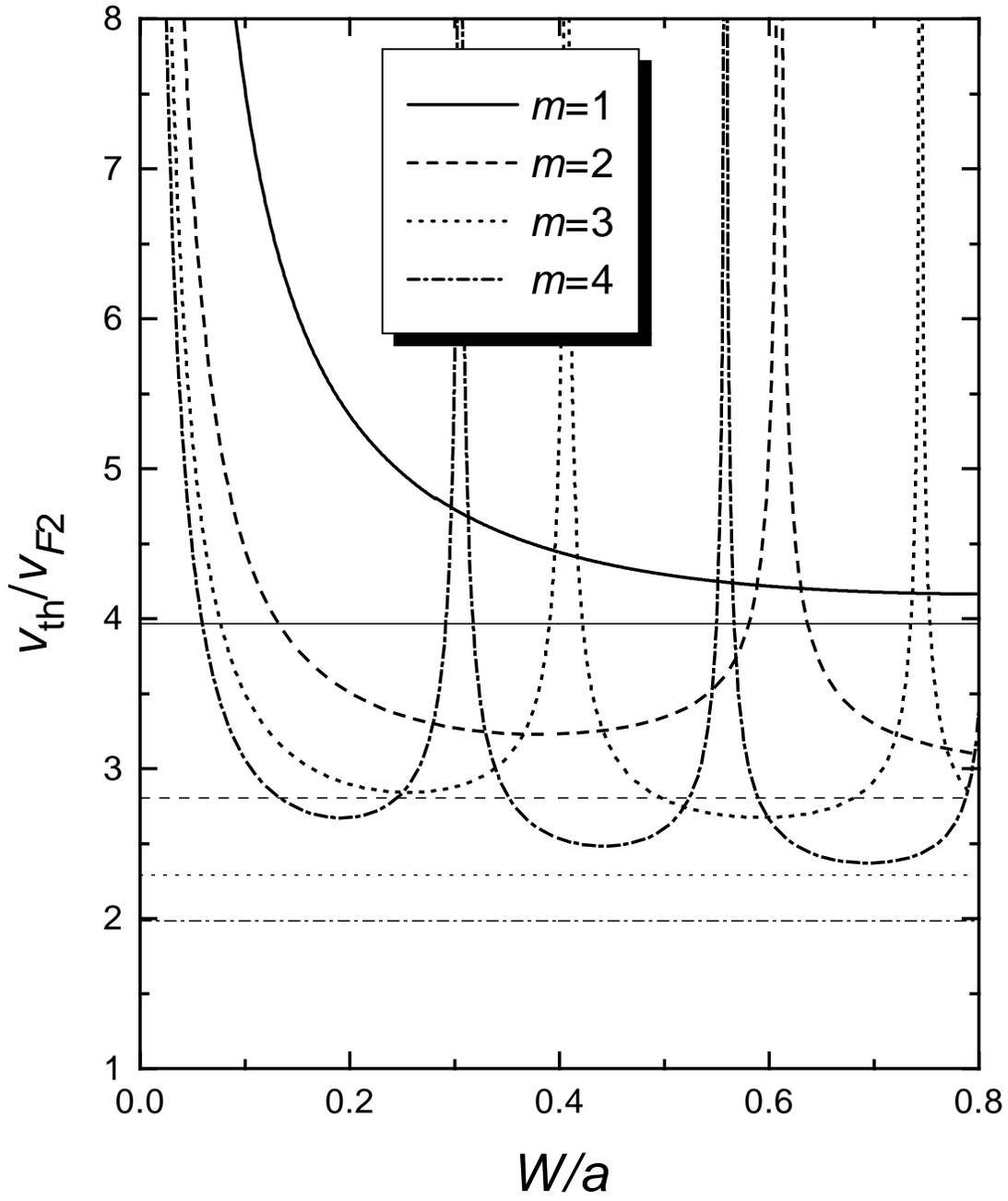
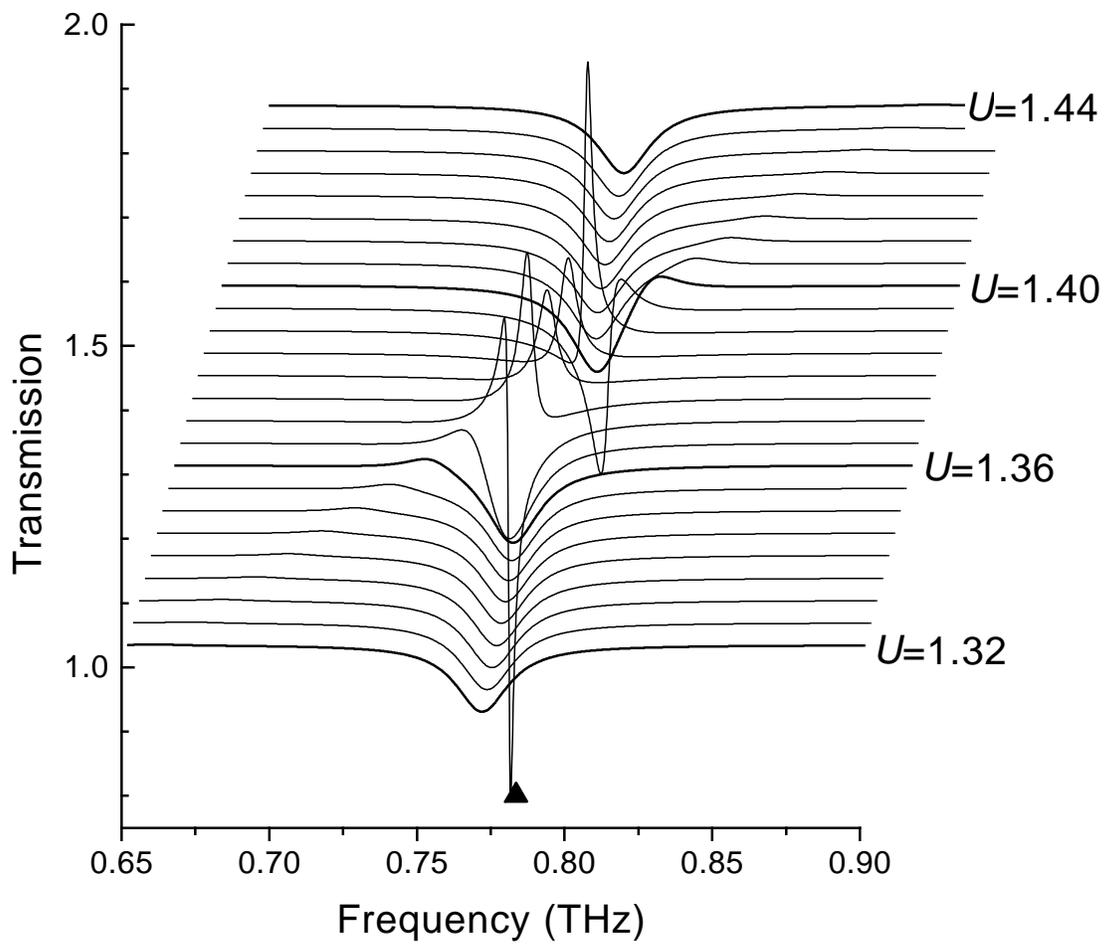
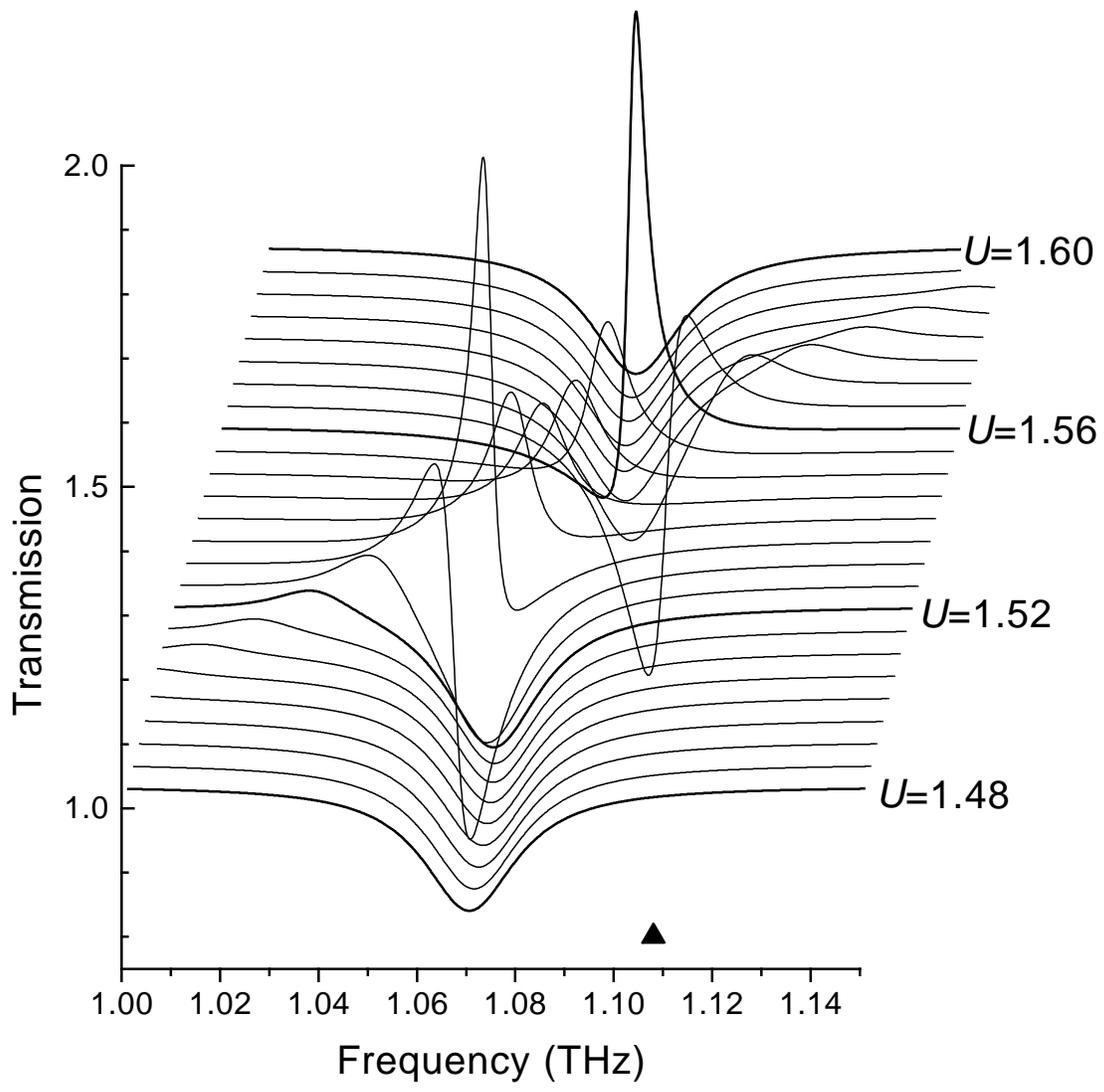
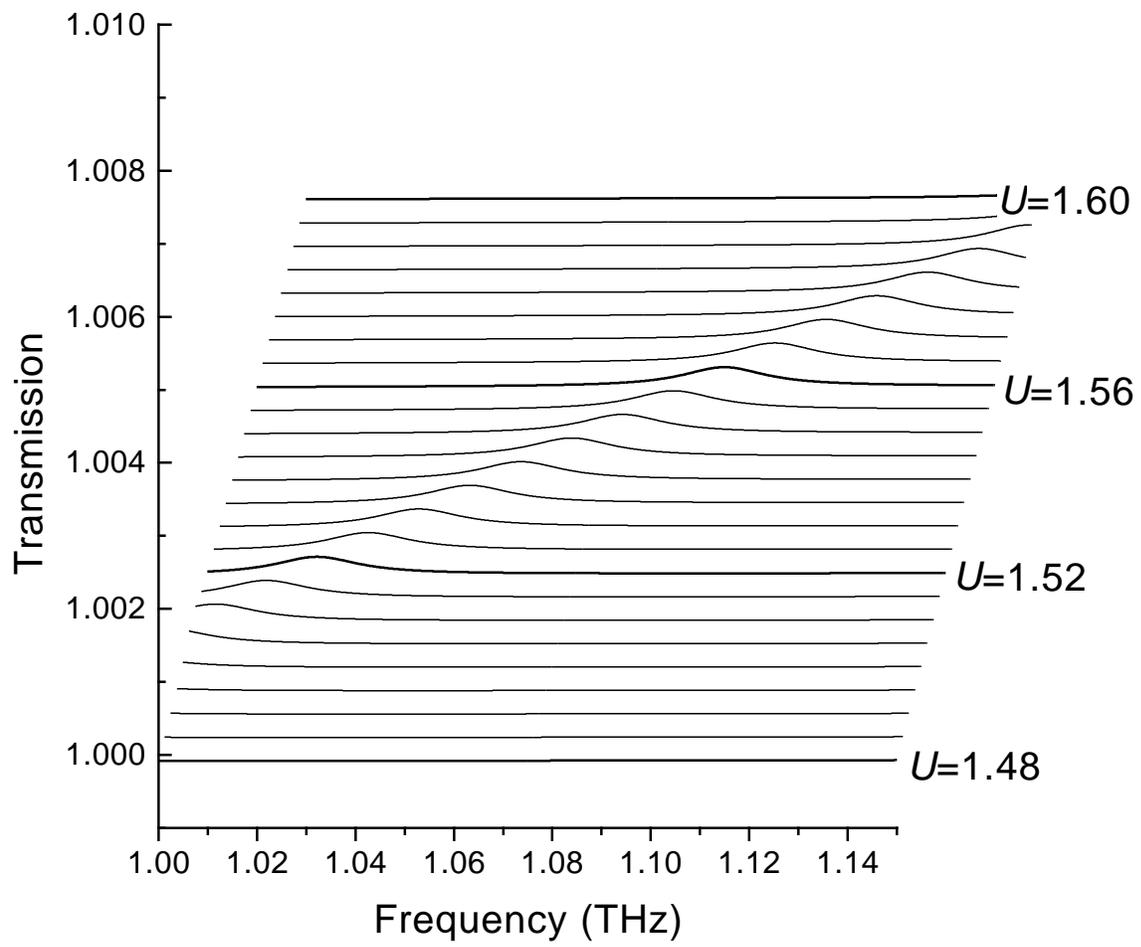
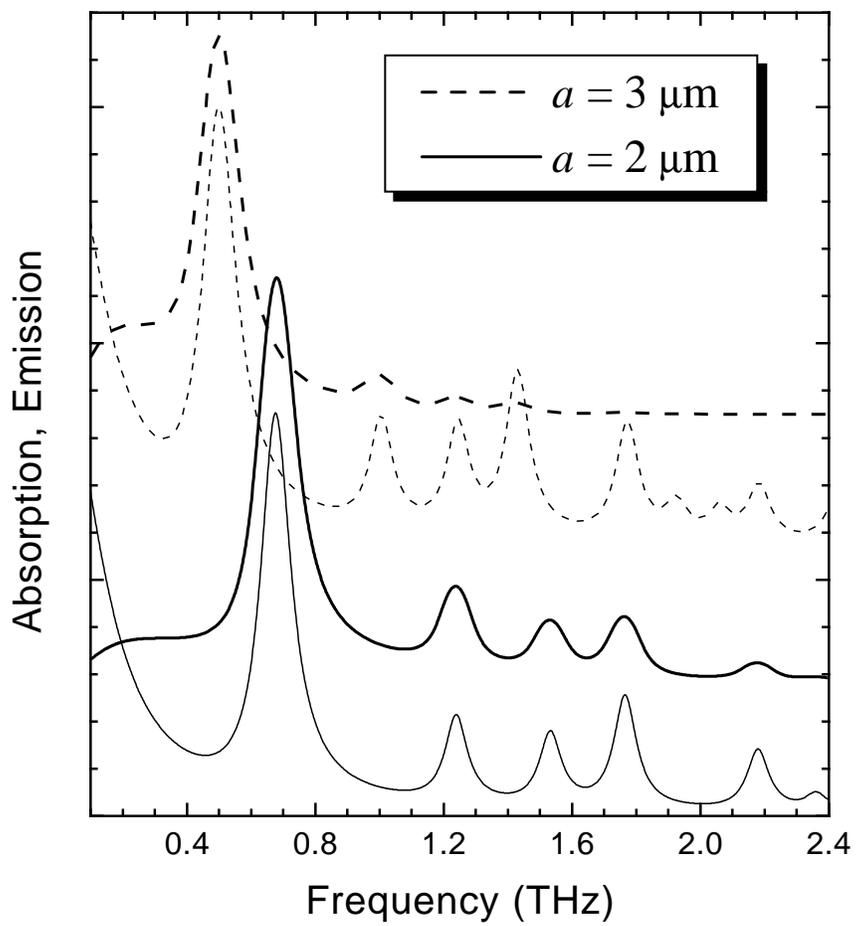
\end{document}